\documentclass{jfm}

\usepackage{amsmath,amssymb,amsfonts}
\usepackage{physics} 

\usepackage[usenames,dvipsnames]{xcolor}
\usepackage{printlen} 

\usepackage{subcaption} 

\usepackage{ragged2e} 

\usepackage{booktabs} 
\usepackage{multirow} 

\usepackage{blindtext} 

\usepackage{tikz}
\usetikzlibrary{arrows.meta} 
\tikzset{obj/.style={fill=cyan!20}}
\usetikzlibrary{patterns, patterns.meta}

\usepackage{graphicx} 
\usepackage{newtxtext}
\usepackage{newtxmath}
\usepackage{natbib}
\usepackage{hyperref}
\hypersetup{
    colorlinks = true,
    urlcolor   = blue,
    citecolor  = blue,
    linkcolor  = blue,
    filecolor  = blue,
}

\newcommand{\RomanNumeralCaps}[1]
\linenumbers

\usepackage{blindtext}

\title{Weakly nonlinear analysis of particle-laden Rayleigh-B\'enard convection}

\author{Thota Srinivas\aff{1},
 \and Gaurav Tomar\aff{1}
 \corresp{\email{gtom@iisc.ac.in}}}

\affiliation{\aff{1}Department of Mechanical Engineering, Indian Institute of Science, Bengaluru-560012, India}

\begin{document}
\maketitle

\begin{abstract}
We investigate the effect of inertial particles on Rayleigh-B\'enard convection using weakly nonlinear stability analysis. In the presence of nonlinear effects, we study the limiting value of growth of instabilities by deriving a cubic Landau equation. An Euler-Euler/two-fluid formulation is being used to describe the flow instabilities in particle-laden Rayleigh-B\'enard convection. The nonlinear results are presented near the critical point (bifurcation point) for water droplets in the dry air system. It is found that supercritical bifurcation is the only type of bifurcation beyond the critical point. Interaction of settling particles with the flow and the Reynolds stress or distortion terms emerge due to the nonlinear self-interaction of fundamental modes, breaking down the top-bottom symmetry of the secondary flow structures. In addition to the distortion functions, the nonlinear interaction of fundamental modes generates higher harmonics, leading to the tendency of preferential concentration of uniformly distributed particles, which is completely absent in the linear stability analysis. It is shown that in the presence of thermal energy coupling between the fluid and particles, the difference between the horizontally averaged heat flux at the hot and cold surface is equal to the net sensible heat flux advected by the particles. The difference between the heat fluxes at hot and cold surfaces is increased with an increase in particle concentration.

\end{abstract}

\begin{keywords}
Key words.
\end{keywords}

{\bf MSC Codes }  {\it(Optional)} Please enter your MSC Codes here

\section{Introduction}\label{sec:introduction}
Particle-laden flows are ubiquitous in nature and industries, such as the dispersion of water droplets in atmospheric clouds \citep{shaw2003particle, abade2018broadening}, aerosols and dust particles in the atmosphere \citep{chandrakar2020droplet}, crystals settling in the earth's magma chambers \citep{martin1988crystal,koyaguchi1990sedimentation, molina2015model}, mixing fuel droplets in the combustion chambers \citep{huang2021eulerian, cao2024effects}, spray coating \citep{pendar2021numerical}, ejection of solid particles and ash due to volcanic eruption \citep{schwaiger2012ash3d,yarushina2015two}, dispersion of pollen grains \citep{helbig2004numerical, robichaud2021numerical}, dynamics of phytoplankton in ocean waters \citep{ruiz2004turbulence,squires1995preferential} etc. Due to this broad spectrum of applications, particle-laden flows are of great interest to fluid dynamists and geophysicists. Hence, to get a comprehensive understanding of these flows, various experimental and numerical techniques have been proposed in the literature \citep{kiger1997dissipation, hwang2006homogeneous, zhong2009enhanced, brandt2022particle,srinivas2025generalized}.

Electronic systems such as teraflop computers, optical fibers, high-energy density lasers, and high-power X-rays generate a lot of heat flux during operation, which makes designing for effective cooling difficult. One way to achieve efficient heat transfer is by using nanofluids, which are known for their higher thermal conductivity and heat transfer coefficients compared to their base fluids \citep{xuan2000heat, maiga2005heat}. \cite{buongiorno2006convective} proposed a homogeneous equilibrium model that provides an explanation for the abnormal increase in the heat transfer coefficient in nanofluids. The reviews of the relevant works can be found in \cite{haddad2012review, javed2020internal, mahian2019recent}. \cite{buongiorno2006convective} showed that the nanoparticles are only sensitive to the processes called Brownian diffusion and thermophoresis. As the particle size increases to the millimetre range and above, these processes quickly become unimportant and are dominated by particle inertia, and the homogeneous equilibrium model developed for the nanofluids is no longer valid. However, few experimental studies have been performed on micrometer-size particles in the literature. For instance, \cite{zhong2009enhanced} performed experiments on droplet condensation of ethane when heated from below and cooled at the top surface below the liquid-vapor coexisting temperature. It is observed that the effective thermal conductivity increases linearly with a decrease in the top surface temperature. This increase in the effective thermal conductivity is an order of magnitude higher than that for the single phase.

\cite{oresta2013effects} studied the effect of settling particles in a weakly turbulent Rayleigh-B\'enard convection. It is shown that the mechanical coupling between the particles and fluid increases the Nusselt number with increasing particle size. However, thermal coupling between the fluid and particles tends to make the fluid temperature uniform and reduce the strength of the convection of the underlying fluid flow. Moreover, \cite{oresta2013effects} reported an unusual kind of reverse one-way coupling in the sense that the underlying flow was affected more significantly than the settling particles, and it is attributed to the continuity equation. Later, \cite{oresta2014multiphase} reviewed the mathematical formulations and the numerical methods for the particle and bubble-laden Rayleigh-B\'enard convection. To understand the settling of dense crystals in magma chambers and planetary scale magma oceans, \cite{patovcka2020settling} conducted a numerical study on the rate of settling of particles in a rectangular two-dimensional Rayleigh-B\'enard convection with Rayleigh number up to $10^{12}$ and Prandtl number $10$ to $50$. Four distinct settling regimes, namely, stone-like, bi-liner, transitional and dust-like regimes, have been observed based on the ratio of particle terminal speed and flow r.m.s velocity. Similarly, \cite{srinivas2024particle} analysed the effect of particle size on the particle cloud patterns in Rayleigh-B\'enard convection. However, in these studies, the coupling between fluid and particles is one-way, in the sense that the fluid flow field affects the particle trajectories while the effect of particles on the underlying flow is ignored. Recently, \cite{denzel2023stochastic} developed a stochastic model to predict the residence times of the particles in a turbulent Rayliegh-B\'enard convection using Euler-Lagrange formulation.

A linear stability analysis of a fluid confined between the more realistic, rigid surfaces and heated from below was performed by Sir Harold Jeffreys \citep{jeffreys1928some} and obtained the critical Rayleigh number $Ra_c$ and critical wave number $k_c$ were obtained as 1708 and 3.117, respectively. However, at present, we know that the collective dynamics of small particles or bubbles can affect the underlying flow significantly through various numerical and experimental studies \citep{hetsroni1971distribution, gore1989effect, sun1986structure, lance1991homogeneous} and the most work has been done for isothermal systems \citep{balachandar2010turbulent, m2016point, maxey2017simulation}. The effect of the suspended particles on the convective heat transfer, particularly Rayleigh-B\'enard convection, is gaining importance in recent times and is the main theme of the current work. \cite{prakhar2021linear} formulated an Euler-Euler or a two-fluid model to study the effect of highly dense point particles on the stability of Rayleigh-B\'enard convection in a horizontally unbounded cell. They reported that the addition of particles into the flow increases the underlying dimensionless parameter space and stabilizes flow significantly. More recently, \cite{raza2024stabilization} extended this stability analysis to the suspension of point bubbles in Rayleigh-B\'enard convection by including additional forces like added mass in the particle momentum balance equations.

 In linear analysis, we initially neglect non-linear perturbation terms that are very small in magnitude; however, as soon as $t=\order{1/c_i}$, they become $\order{1}$ in magnitude and cannot be neglected, here, $c_i$ is the growth rate. In other words, as the Rayleigh number $Ra>Ra_c$, the initial infinitesimally small perturbations grow exponentially and reach a magnitude and can affect the mean flow, which makes the linear stability predictions unreliable. Indeed, the non-linear terms might quench the exponential growth and lead to a steady or oscillating solution for ${Ra}$ slightly above ${Ra}_c$ \citep{cross1993pattern,cross2009pattern}. More generally, as \cite{hof2006finite} mentioned, with an increase in the flow velocity, the transition from the smooth laminar to the highly disordered turbulent flow can occur through a series of instabilities during which the system encounters progressively complicated states \citep{niemela2000turbulent}, or it occurs abruptly \citep{grossmann2000onset, hof2004experimental}. Out of these two routes, \cite{busse2003sequence} described a sequence of bifurcations to complex fluid flow to occur from the simple laminar flow. Beyond the onset of convection, linear stability, in general, can not predict the nature of instabilities and secondary flow patterns that occur in the flow field. The temporal evolution of the perturbation amplitude can be studied using the nonlinear analysis. There are three approaches available for nonlinear analysis: (i) weakly nonlinear stability analysis, (ii) Direct Numerical Simulations (DNS), and (iii) deflation technique. Near the onset of convection, weakly nonlinear stability analysis gives valuable insights into the nature of instabilities and the secondary flow patterns with minimum computational cost compared to DNS. The deflation technique is a recent method devised to obtain the nontrivial distinct solutions of nonlinear partial differential equations \cite{farrell2015deflation, farrell2016computation}. Using this technique, \cite{boulle2022bifurcation} bifurcation analysis of steady states of two-dimensional Rayleigh-B\'enard convection with no-slip boundary conditions. 
 
 In the current work, we use the Euler-Euler formulation given by \cite{prakhar2021linear} and perform a weakly nonlinear stability analysis to study the effect of particles on the underlying flow beyond the onset of convection near the critical point. As the perturbation amplitude grows, the nonlinear terms become significant and might lead to the phenomenon of preferential concentration of particles, which is completely absent in the linear stability analysis. Hence, we study the effect of nonlinear terms on the secondary flow patterns beyond the onset of convection.

The rest of the paper is organized as follows. The problem statement and the mathematical model are given in \S\ref{sec:model}. In \S\ref{sec:linear stability}, the linear stability analysis is provided, together with the governing equations of the basic flow and linear disturbances and a brief review of the linear stability findings. The full formulation of the amplitude equation using weakly nonlinear stability analysis and the analysis of mutual energy exchange between fluid and particles is provided in \S\ref{sec:amplitude equation formulation}. The numerical procedure for the present work is shown in appendix \ref{appen:numerical procedure}. The results and discussion are presented in \S\ref{sec:results and discussion}. Lastly, a brief summary of the current study is given in \S \ref{sec:conclusions}.

\section{Problem statement and mathematical model}\label{sec:model}
\begin{figure}
    \centering
    \resizebox {0.70\textwidth} {!} {
        \begin{tikzpicture}[>=stealth]
        \draw[thick,dashed] (0,0)--++(0,4.9927);
        \draw[thick,dashed] (8,0)--++(0,4.9927);
        \draw[thick] (0,0)--++(8,0);
        \draw[thick] (0,4.9927)--++(8,0);

        \draw[pattern={north east lines},pattern color=black]
        (0,0) rectangle +(8,0.1);
        \draw[pattern={north east lines},pattern color=black]
        (0,4.9927) rectangle +(8,-0.1); 

        \filldraw[style={fill=cyan!20}] (0,0.1) rectangle (8,4.9927-0.1);
        
        \draw[pattern={Dots[angle=45,distance={2pt/sqrt(2)},radius=0.2pt]}, pattern color=red]
        (0,0.1) rectangle +(8,4.8);

        \draw[thin] (7,3.9) circle (0.2);
        \draw[thin] (7,4.1) --++(2,0.5);
        \draw[thin] (7,3.7) --++(2,-0.5);
        \filldraw[thick, style={fill=cyan!20}] (9.1,3.9) circle (0.7);
        \draw[pattern={Dots[angle=45,distance={6pt/sqrt(2)},radius=0.8pt]}, pattern color=red]
        (9.1,3.9) circle (0.7);
        \node at (9.1,4.8) {particles};
        
        \draw[thick,->] (0,2.49635) -- ++ (1.5,0);
        \draw[thick,->] (0,2.49635) -- ++ (0,1.5);
        \node at (1.5,2.29635) {$x$};
        \node at (-0.2,3.99635) {$z$};
        \node at (-1.1,5.) {$z^*=+H/2$};
        \node at (-1.1,-0.05) {$z^*=-H/2$};
        \node at (-0.2,2.49635-0.1) {O};
        \node at (4,-0.4) {$T^*=T_h$, $u_x^*=u_z^*=0$};
        \node at (4,5.25) {$T^*=T_c$, $\vb{u}^*=0$, $\vb{v}^*=-v_\tau \vu{e}_z$, $T_p^*=T_{pt}$, $\phi=\Phi_0$};
\end{tikzpicture}
    }
    \caption{Schematic of the particle-laden Rayleigh--B\'enard convection.}
    \label{fig:prbc schematic}
\end{figure}
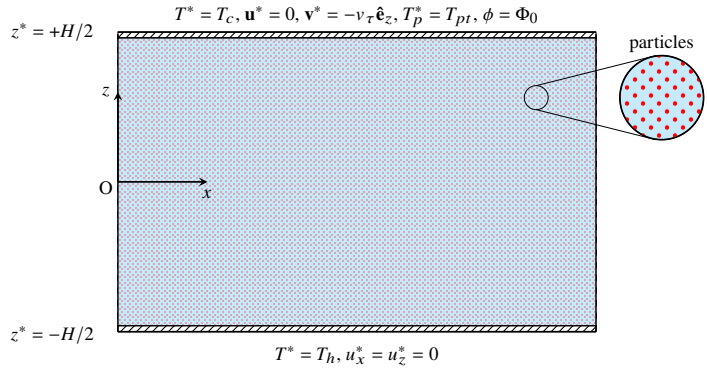

The physical model and domain with schematic representation, as shown in figure \ref{fig:prbc schematic}, consists of continuous fluid domain, $\Omega=\qty{\qty(x^*,\,z^*)\in\mathbb{R}\times\qty(-H/2,\, H/2)}$ along with dispersed particulate phase. Here, the bottom and the top surfaces are treated as isothermal walls maintained at temperatures $T_h$ and $T_c$, respectively. The present study considers the mono-dispersion of very tiny spherical particles at relatively small volume fractions ($\leq 10^{-3}$). There are two alternative methods, called Euler-Euler and Euler-Lagrange formulations, available for the particles in the millimetre range for which the homogeneous equilibrium model fails. However, the choice of a particular method depends on various length scales, such as particle size $d_p$, the smallest flow length scale $\Delta x$, inter-particle separation $\lambda$, and particle number density $n$. When $d_p\ll\Delta\sim\lambda$ and $n\gg1$, the Euler-Lagrange formulation is preferred, whereas if $d_p\ll\lambda\ll\Delta$ and $n\gg1$, the Euler-Euler also called as the two-fluid formulation is preferred. We adopt the two-fluid model given by \citet{prakhar2021linear}, in which the particles are introduced steadily and uniformly at the top surface at their terminal velocity with a fixed temperature. The momentum equation for the particulate phase is given by the volume-averaged Maxey-Riley-Gatignol equation \citep{maxey1983equation, gatignol1983faxen} and the thermal energy equation ensures the balance between the rate of change of the sensible heat of the particles and the rate of convective heat transfer between particulate and fluid. The dimensional independent variables $\left(x^*,\,z^*,\,t^*\right)$ and dependent variables $\left(u_x^*,\,u_z^*,\,p^*,\,T^*,\,v_x^*,\,v_z^*,\,T_p^*\right)$ are non-dimensionalized as follows:
\begin{align}
   \left(x,\,z,\,t\right)&=\left(x^*/H,\,z^*/H,\,t^*U/H\right),\\
   \left(u_x,\,u_z,\,p,\,\theta\right)&=\left(u_x^*/U,\,u_z^*/U,\,p^*/(\rho_fU^2),\,(T-T_c)/\Delta T\right),\\
     \left(v_x,\,v_z,\,\theta_p\right)&=\left(v_x^*/U,\,v_z^*/U,\,(T_p-T_c)/\Delta T\right),
\end{align}
where $x$, $z$, and $t$ are the non-dimensional horizontal coordinate, vertical coordinate and time, respectively. Furthermore, $u_x$, $u_z$, $p$, and $\theta$ are the fluid horizontal velocity component, vertical velocity component, and pressure, respectively. Similarly, $v_x$, $v_z$, and $\theta_p$ are dimensionless particulate phase horizontal velocity component, vertical velocity component and temperature, respectively. Here, we use the distance between the two horizontal surfaces $H$ as a length scale, fluid free-fall velocity $U=\sqrt{g\beta\Delta TH}$ as a velocity scale, and the temperature difference $\Delta T=T_h-T_c$ as a temperature scale. Where $\rho_f$, $g$, and $\beta$ are fluid density, the acceleration due to gravity, and the volume expansion coefficient of fluid, respectively. The resulting governing equations in non-dimensional form are given as

\begin{eqnarray}\label{eq:fluid mass balance}
    \pdv{u_x}{x}&+&\pdv{u_z}{z}=0,\\ \label{eq:ux equation}
    \pdv{u_x}{t}+u_x\pdv{u_x}{x}+u_z\pdv{u_x}{z}&=&-\pdv{p}{x} + \sqrt{\frac{{\Pran}}{{Ra}}}\nabla^{2}u_x-\frac{{R}\phi}{{St}_m}\qty(u_x-v_x),\\ \label{eq:uz equation}
    \pdv{u_z}{t}+u_x\pdv{u_z}{x}+u_z\pdv{u_z}{z}&=&-\pdv{p}{z} + \sqrt{\frac{{\Pran}}{{Ra}}}\nabla^{2}u_z+\theta-\frac{{R}\phi}{{St}_m}\qty(u_z-v_z),\\ \label{eq:fluid energy equation}
    \pdv{\theta}{t}+u_x\pdv{\theta}{x}+u_z\pdv{\theta}{z}&=&\frac{1}{\sqrt{{Ra\Pran}}}\nabla^2 \theta-\frac{{E}\phi}{{St}_{th}}\qty(\theta-\theta_p),\\
    \pdv{\phi}{t}+\pdv{\qty(\phi v_x)}{x}&+&\pdv{\left(\phi v_z\right)}{z}=0,\\
    \pdv{v_x}{t}+v_x\pdv{v_x}{x}&+&v_z\pdv{v_x}{z}=\frac{u_x-v_x}{{St}_m},\\
    \pdv{v_z}{t}+v_x\pdv{v_z}{x}&+&v_z\pdv{v_z}{z}=\frac{u_z-v_z}{{St}_m}-\frac{v_0}{{St}_m},\\ \label{eq:particle energy balance}
    \pdv{\theta_p}{t}+v_x\pdv{\theta_p}{x}&+&v_z\pdv{\theta_p}{z}=\frac{\theta-\theta_p}{{St}_{th}}
\end{eqnarray}
for all $\left(x,\,z,\,t\right)\in\left\{\mathbb{R}\times\left(-1/2,\, 1/2\right)\times\left(0,\, \infty\right)\right\}$. 
The boundary conditions are as follows:
\begin{eqnarray}
        \mbox{At }z&=&1/2:\mbox{\ }u_x=u_z=0,\mbox{ }\theta=0,\mbox{ }v_x=0,\mbox{ }v_z=-v_0,\mbox{ }\theta_p=\Theta_{pt},\mbox{ and }\phi=\Phi_0\\
        \mbox{At }z&=&-1/2:\mbox{\ }u_x=u_z=0,\mbox{ and }\theta=1
\end{eqnarray}
for all $\qty(x,\,t)\in\left\{\mathbb{R}\times\qty(0,\,\infty)\right\}$. Where $\phi$ is the particle volume fraction field and the non-dimensional parameters present in the problem, Rayleigh number $(Ra)$, Prandtl number $(\Pran)$, density ratio $(R)$, specific heat capacity ratio $(E)$, mechanical Stokes number $(St_m)$, thermal Stokes number $(St_{th})$, initial particles temperature $(\Theta_{pt})$, and non-dimensional particle terminal speed $(v_0)$ are defined as:

\refstepcounter{equation}
$$
  Ra=g\beta\Delta TH^3/(\alpha_f\nu_f), \quad
  \Pran=\nu_f/\alpha_f, \quad
  R=\rho_p/\rho_f,
$$
$$
 {St}_m=\frac{\tau_p}{\tau_f}=\frac{{R}\delta^2}{18{C}_d} \sqrt{\frac{{Ra}}{{\Pran}}},\quad
 {St}_{th}=\frac{\tau_{th}}{\tau_f}=\frac{{E}\delta^2}{6{Nu}_p}\sqrt{{Ra\Pran}},
  \eqno{(\theequation{\mathit{a}-\mathit{h}})}\label{eq:dimensionless parameters}
$$
$$
 E=R(C_{pp}/C_{pf}), \quad
 v_0=\frac{{\Rey}_p}{\delta}\sqrt{\frac{{\Pran}}{{Ra}}}, \quad
\Theta_{pt}=\frac{T_{pt}-T_c}{T_h-T_c},
$$
where $\alpha_f=\kappa_f/\rho_fC_{pf}$ is the fluid thermal diffusion coefficient, $\nu_f=\mu_f/\rho_f$ is the fluid kinematic viscosity, $\kappa_f$ is the fluid thermal conductivity, $C_{pf}$ is the fluid specific heat constant, $\mu_f$ is fluid dynamic viscosity, $\rho_p$ is the particle density, and $C_{pp}$ is the particle specific heat. Particle diameter $d_p$ in the non-dimensional form is represented as $\delta=d_p/H$. Introduction of particles into the flow leads to three distinct time scales, namely, $\tau_f=H/U$ flow time scale, $\tau_p=\rho_pd_p^2/(18\mu_fC_d(\Rey_p))$ particle mechanical relaxation time scale, and $\tau_{th}=\rho_pC_{pp}d_p^2/(6\kappa_fNu_p(\Rey_p,\, \Pran))$ is the particle thermal relaxation time scale. Here ${C}_d=1+0.15{\Rey}_p^{0.687}$ is the Schiller-Naumann correction factor to the Stokes drag when the particle Reynolds number ${\Rey}_p=\rho_fv_\tau d_p/\mu_f$ is greater than unity and less than 800 \citep{clift2005bubbles}, $v_\tau=\qty(1-\rho_f/\rho_p)\tau_pg$ is the settling speed of a particle in the quiescent fluid, and $Nu_p=2.0+0.6\Rey_p^{1/2}\Pran^{1/3}$ is the particle Nusselt number. The last terms on the right-hand side of equations (\ref{eq:ux equation}) and (\ref{eq:uz equation}) represent the mechanical two-way coupling between the fluid and particles, whereas the thermal two-way coupling is captured by the right-hand side last term of (\ref{eq:fluid energy equation}). 

\section{Linear stability analysis}\label{sec:linear stability}
We assume the fluid-particle system is at a steady state with particles settling uniformly with their terminal velocity in a quiescent fluid. Under these conditions, the above governing equations (\ref{eq:fluid mass balance})--(\ref{eq:particle energy balance}) reduced to a system of ordinary differential equations, which are represented in the operator form as

\begin{eqnarray}\label{eq:basestate uz equation}
        \mathcal{L}_z\qty(P_0,\,\Theta_0,\,v_0,\,\Phi_0,\,{R},\,{St}_m)&=&\dv{P_0}{z}-\Theta_0+\frac{{R}\Phi_0}{{St}_m}v_0=0,\\ \label{eq:basestate theta equation}
        \mathcal{L}_\theta\qty(\Theta_0,\,\Theta_{p0},\,\Phi_0,\,{Ra},\,{\Pran},\,{E})&=&\frac{1}{\sqrt{{Ra\Pran}}}\dv[2]{\Theta_0}{z}-\frac{{E}\Phi_0}{{St}_{th}}\qty(\Theta_0-\Theta_{p0})=0,\\ \label{eq:basestate thetap equation}
        \mathcal{L}_{p\theta}\qty(\Theta_0,\,\Theta_{p0},\,v_0,\,{St}_{th})&=&\frac{\Theta_0-\Theta_{p0}}{{St}_{th}}+v_0\dv{\Theta_{p0}}{z}=0
\end{eqnarray}
for $z\in\qty(-1/2,\,1/2)$ with boundary conditions
\begin{align}
    \text{At }z=&1/2:\text{ }\Theta_0=0,\text{ and } \Theta_{p0}=\Theta_{pt},\\
    \text{At }z=&-1/2:\text{ }\Theta_0=1,
\end{align}
where $P_0$, $\Theta_0$, and $\Theta_{p0}$ are the basic state fluid pressure, basic state fluid temperature, and basic state particle temperature field, respectively, and $\Phi_0$ is the initial uniform particle volume fraction. Using the above boundary conditions, the solution to the equations (\ref{eq:basestate uz equation})-(\ref{eq:basestate thetap equation}) yields the base state given by,
\begin{eqnarray}
    \Theta_0&=&\Theta_{pt}+\mathscr{A}\qty[1-\qty(1-lm_1)e^{-m_1(1/2-z)}]+\mathscr{B}\qty[1-\qty(1-lm_2)e^{-m_2(1/2-z)}],\\
    \Theta_{p0}&=&\Theta_{pt}+\mathscr{A}\qty[1-e^{-m_1(1/2-z)}]+\mathscr{B}\qty[1-e^{-m_2(1/2-z)}],\\
    \dv{P_0}{z}&=&\Theta_{pt}+\mathscr{A}\qty[1-\qty(1-lm_1)e^{-m_1(1/2-z)}]+\mathscr{B}\qty[1-\qty(1-lm_2)e^{-m_2(1/2-z)}]\nonumber\\
    &&\quad-\frac{{R}\Phi_0}{{St}_m}v_0,
\end{eqnarray}
where $\mathscr{A}$ and $\mathscr{B}$ are integration constants given by,
\begin{eqnarray}
    \mathscr{A}&=&-\frac{(1-\Theta_{pt})lm_2+\Theta_{pt}\qty[1-\qty(1-lm_2)e^{-m_2}]}{lm_1\qty[1-\qty(1-lm_2)e^{-m_2}]-lm_2\qty[1-\qty(1-lm_1)e^{-m_1}]},\\
    \mathscr{B}&=&+\frac{(1-\Theta_{pt})lm_1+\Theta_{pt}\qty[1-\qty(1-lm_1)e^{-m_1}]}{lm_1\qty[1-\qty(1-lm_2)e^{-m_2}]-lm_2\qty[1-\qty(1-lm_1)e^{-m_1}]},
\end{eqnarray}
here, $m_1$ and $m_2$ are given by
\refstepcounter{equation}
$$
  m_1=\frac{1}{2l}\qty[1+\sqrt{1+4\qty(\frac{l}{{L}})^2}],\quad
  m_2=\frac{1}{2l}\qty[1-\sqrt{1+4\qty(\frac{l}{{L}})^2}],\eqno{(\theequation{\mathit{a},\,\mathit{b}})}\label{eq:m1 and m2}
$$
where ${L}$ is the spatial non-dimensional length scale over which the effect of a particle on the surrounding fluid temperature is significant \citep{prakhar2021linear} and is given by
\begin{equation}\label{eq:L definition}
    {L}=\qty(\frac{{St}_{th}}{{E}\Phi_0\sqrt{{Ra\Pran}}})^{1/2}=\frac{\delta}{\sqrt{6\Phi_0{Nu}_p}},
\end{equation}
where $l=v_0{St}_{th}$ is the non-dimensional length scale that the particle must traverse for its temperature to be locally equal to that of the fluid \citep{prakhar2021linear} and is related to other parameters as follows
\begin{equation}\label{eq:l definition}
    l=\frac{{St}_{th}{\Rey}_p}{\delta}\sqrt{\frac{{Pr}}{{Ra}}}=\frac{{E}\delta{\Pran}{\Rey}_p}{6{Nu}_p}.
\end{equation}
The classical normal mode analysis \citep{drazin2004hydrodynamic} is performed to examine the stability of the basic flow mentioned above. In general, the linear stability analysis is carried out by decomposing all the dependent variables into a steady basic state and the respective infinitesimal perturbations (represented by a superscript prime),
\begin{equation}\label{eq:base + perturbations PRBC}
    \begin{split}
    \qty(u_x,\,u_z,\,\theta,\,v_x,\,v_z,\,\theta_p,\,p,\,\phi)=\qty(0,\,0,\,\Theta_0(z),\,0,\,-v_0,\,\Theta_{p0}(z),\,P_{0}(z),\,\Phi_0)\\
    +\qty(u_x',\,u_z',\,\theta',\,v_x',\,v_z',\,\theta_p',\,p',\,\phi'),
    \end{split}
\end{equation}
After neglecting the higher order and retaining only the first order terms in\\ $\qty{u_x',\,u_z',\,\theta',\,v_x',\,v_z',\,\theta_p',\,p',\,\phi'}$, we obtain,
\begin{eqnarray}\label{eq:linear fluid masss balance}
        \pdv{u_x'}{x}&+&\pdv{u_z'}{z}=0,\\
        \pdv{u_x'}{t}&=&-\pdv{p'}{x}+\sqrt{\frac{{\Pran}}{{Ra}}}\laplacian{u_x'}-\frac{{R}\Phi_0}{{St}_m}\qty(u_x'-v_x'),\\
        \pdv{u_z'}{t}&=&-\pdv{p'}{z}+\sqrt{\frac{{\Pran}}{{Ra}}}\laplacian{u_z'}+\theta'-\frac{{R}\Phi_0}{{St}_m}\qty(u_z'-v_z')-\frac{{R}v_0}{{St}_m}\phi',\\
        \pdv{\theta'}{t}&=&\frac{1}{\sqrt{{Ra\Pran}}}\laplacian\theta'-u_z'\dv{\Theta_0}{z}-\frac{{E}\Phi_0}{{St}_{th}}\qty(\theta'-\theta_p')-\frac{{E}\qty(\Theta_0-\Theta_{p0})}{{St}_{th}}\phi',\\
        \pdv{\phi'}{t}&=&v_0\pdv{\phi'}{z}-\Phi_0\qty(\pdv{v_x'}{x}+\pdv{v_z'}{z}),\\
        \pdv{v_x'}{t}&=&v_0\pdv{v_x'}{z}+\frac{u_x'-v_x'}{{St}_m},\\
        \pdv{v_z'}{t}&=&v_0\pdv{v_z'}{z}+\frac{u_z'-v_z'}{{St}_m},\\ \label{eq:linear partilce energy}
        \pdv{\theta_p'}{t}&=&v_0\pdv{\theta_p'}{z}+\frac{\theta'-\theta_p'}{{St}_{th}},
\end{eqnarray}
The above system of equations is linear homogeneous and has the coefficients functions of spatial variables only, but not dependent on time. Hence, they satisfy the solution in the normal mode form given by
\begin{equation}\label{eq:normal modes H2T2}
    \begin{split}
    &\qty(u_x',\,u_z',\,\theta',\,v_x',\,v_z',\,\theta_p',\,p',\,\phi')^T\\
    &\qquad=e^{\displaystyle\mbox{i}k\qty(x-ct)}\qty(\hat{u}_x(z),\,\hat{u}_z(z),\,\hat{\theta}(z),\,\hat{v}_x(z),\,\hat{v}_z(z),\,\hat{\theta}_p(z),\,\hat{p}(z),\,\hat{\phi}(z))^T
    \end{split}
\end{equation}
where $k$ is wave number and $c=c_r+\text{i}c_r$ is the complex wave speed corresponding to $k$. The sign of $c_i$ determines the growth or decay of the disturbance. That is, as $c_i<0$ or $c_i=0$ or $c_i>0$, the flow is stable or neutrally stable or unstable, respectively.
Substitution of equations (\ref{eq:normal modes H2T2}) in equations (\ref{eq:linear fluid masss balance})--(\ref{eq:linear partilce energy}) leads to linearized disturbance equations in operator form given in the appendix \ref{appen:linearized disturbance equations}. The boundary conditions for corresponding equations are given by
\begin{eqnarray}\label{eq:linear bc at z=1/2}
    \text{At }z&=&1/2:\text{ }\hat{u}_x=\hat{u}_z=\hat{\theta}=\hat{v}_x=\hat{v}_z=\hat{\theta}_p=\hat{\phi}=0,\\ \label{eq:linear bc at z=-1/2}
    \text{At }z&=&-1/2:\text{ }\hat{u}_x=\hat{u}_z=\hat{\theta}=0.
\end{eqnarray}
The system of linear equations (\ref{eq:linear continuity equation})--(\ref{eq:thetap linear equation}) together with the boundary conditions gives rise to a generalized eigenvalue problem given by
\begin{equation}\label{eq:gen. eigenvalue system for h2t2}
    \mathbb{A}\vb{q}=c\mathbb{B}\vb{q},
\end{equation}
where $\vb{q}=\qty(\hat{u}_x,\,\hat{u}_z,\,\hat{\theta},\,\hat{v}_x,\,\hat{v}_z,\,\hat{\theta}_p,\,\hat{p},\,\hat{\phi})$ is the eigenfunction corresponding to the eigenvalue $c$ and $\qty(\mathbb{A},\,\mathbb{B})$ are the square complex matrices. The fundamental disturbance is given by $\vb{q}e^{\displaystyle\mbox{i}k\qty(x-c_rt)}$, here $\vb{q}$ is eigenfunction related to the least stable eigenvalue and $c_r$ is the corresponding wave speed (Several parts of the weakly nonlinear analysis will employ the frequent use of the fundamental disturbance). 

The bifurcation point, also known as the critical point $(k, Ra)$, and the shape of the emerging disturbances are both determined using the linear stability theory. It offers no information regarding the actual magnitude of the disturbances (amplitude) away from the critical value. A weakly nonlinear stability study is necessary to examine the amplitude of such disturbances. The outcomes of the linear stability analysis are also necessary for a nonlinear analysis.

There have been two studies \citep{prakhar2021linear, raza2024stabilization} on the linear stability analysis of the present problem in the literature. Using a two-fluid model \cite{prakhar2021linear} studied the stability threshold of particle-laden Rayleigh-B\'enard convection when the particles are much denser than the underlying fluid. \cite{raza2024stabilization} extended the two-fluid model by \cite{prakhar2021linear} to the lighter particles like bubbles by adding added-mass term. Using dimensional analysis, we can show that eight independent dimensionless parameters exist in this problem such as Rayleigh number $(Ra)$, Prandtl number $(\Pran)$, density ratio $(R)$, heat capacity ratio $(E)$, particle Reynolds number $(\Rey_p)$, particle diameter $\delta$, particle injection temperature $(\Theta_{pt})$, and initial particle volume fraction $\Phi_0$. We fix the dimensionless parameters such as Prandtl number $\Pran=0.71$, density ratio $R=800$, and specific heat capacity ratio $E=3385$, such that the fluid-particle system represents the water droplets suspended in dry air. 

\begin{figure}
    \centering
    \begin{subfigure}[b]{0.49\textwidth}
        \includegraphics[width=\textwidth]{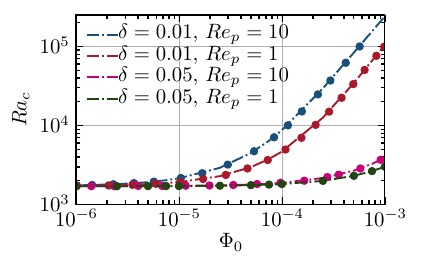}
        \caption{}
        \label{fig:rac vs phi0 h2t2}
    \end{subfigure}
    \hfill
    \begin{subfigure}[b]{0.49\textwidth}
        \includegraphics[width=\textwidth]{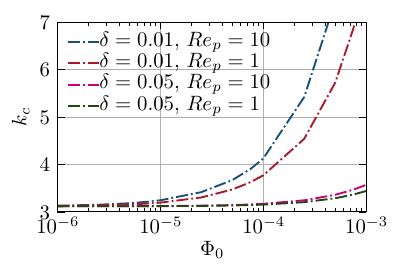}
        \caption{}
        \label{fig:kc vs phi0 h2t2}
    \end{subfigure}
    \captionsetup{width=\linewidth}
    \caption{\justifying Effect of initial particle volume fraction on critical Rayleigh number $Ra_c$ and critical wave number $k_c$: a) variation in critical Rayleigh number ${Ra}_c$, and b) variation in critical wave number $k_c$ with initial undisturbed particle volume fraction $\Phi_0$ for two different particles Reynolds numbers ${\Rey}_p$, and the particle sizes $\delta$, and other parameters kept at $\Theta_{pt}=0$, ${R}=800$, ${E}=3385$, and ${\Pran}=0.71$ for both graphs. Here, the circles represent the data from \cite{prakhar2021linear}.}
    \label{fig:phi vs rac and kc h2t2}
\end{figure}

We present the linear stability results for the air-water droplet system. The effect of initial particle volume fraction on the critical Rayleigh number ${Ra}_c$ and the critical wave number $k_c$ for different particle Reynolds numbers ${\Rey}_p$ and the particle sizes $\delta$ is shown in figure \ref{fig:phi vs rac and kc h2t2}(a,b). The critical Rayleigh number increases with $\Phi_0$ for all ${\Rey}_p$ and $\delta$. This stabilization of the flow under the presence of particles is due to the dissipative nature of the mechanical and thermal two-way coupling source terms in equations (\ref{eq:ux equation})-(\ref{eq:uz equation}), and equations (\ref{eq:fluid energy equation}), respectively. The strength of the mechanical source terms is proportional to $({R}\Phi_0/{St}_m)\sim\Phi_0{C}_d/\delta^2$ independent of the density ratio ${R}$. Similarly, the strength of the thermal source term is proportional to ${E}\Phi_0/{St}_{th}\sim\Phi_0{Nu}_p/\delta^2$ which is independent of the heat capacity ratio ${E}$. Hence, with an increase in the $\Phi_0$ value (decrease in $\delta$ value) in either source terms, the dissipation increases and thereby increases the stability (increase in ${Ra}_c$ value). However, the increase in ${Ra}_c$ with $\Phi_0$ is more significant for smaller particles ($\delta=0.01$) due to the strong $1/\delta^2$ dependence of the two-way coupling source terms than for the large particles ($\delta=0.05$). Moreover, the weak dependency of the mechanical and thermal source terms on particle Reynolds number $Re_p$ through drag force coefficient ${C}_d$ and the Nusselt number $Nu_p$ explains the small increase in ${Ra}_c$ with the increase in ${\Rey}_p$. Similarly, from the figure \ref{fig:kc vs phi0 h2t2}, it is clear that the critical wave number $k_c$ increases with an increase in $\Phi_0$ for all ${\Rey}_p$ and $\delta$. The plausible explanation for this can be obtained using the inter-particle distance $\lambda\sim(1-\Phi_0)^{1/3}\delta/\Phi_0^{1/3}\approx\delta/\Phi_0^{1/3}$ for the dilute suspensions $(\Phi_0\ll1)$ \citep{prakhar2021linear}. Hence, the critical wave number can be expected to vary as $k_c\sim{\Phi_0^{1/3}}/{\delta}$ which explains the increase of $k_c$ with an increase in $\Phi_0$ for all ${\Rey}_p$ and $\delta$ values. We note that the increase in $k_c$ is significant for the small particles $(\delta=0.01)$ with small Reynolds number ${\Rey}_p=1$ than for the large particles $(\delta=0.05)$ with large Reynolds number $({\Rey}_p=10)$ as shown in figure \ref{fig:kc vs phi0 h2t2}.

\begin{figure}
    \centering
    \begin{subfigure}[b]{0.49\textwidth}
        \includegraphics[width=\textwidth]{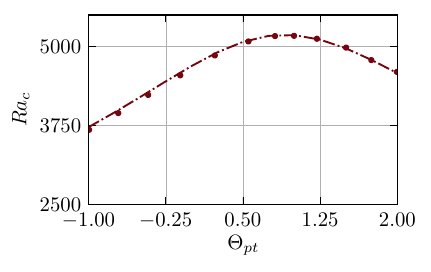}
        \caption{}
        \label{fig:tps vs rac h2t2}
    \end{subfigure}
    \hfill
    \begin{subfigure}[b]{0.49\textwidth}
        \includegraphics[width=\textwidth]{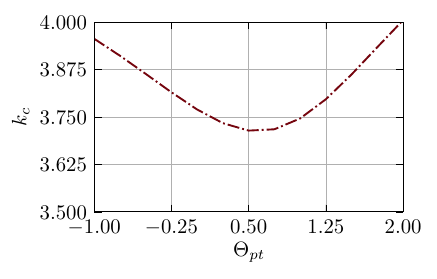}
        \caption{}
        \label{fig:tps vs kc h2t2}
    \end{subfigure}
    \hfill
    \begin{subfigure}[b]{0.51\textwidth}
        \includegraphics[width=\textwidth]{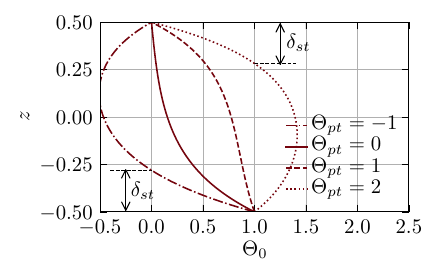}
        \caption{}
        \label{fig:basestate Theta0 h2t2}
    \end{subfigure}
    \hfill
    \begin{subfigure}[b]{0.46\textwidth}
        \includegraphics[width=\textwidth]{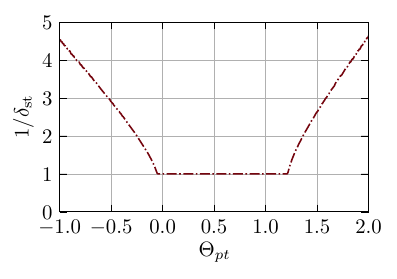}
        \caption{}
        \label{fig:basestate stratification thickness h2t2}
    \end{subfigure}
    \captionsetup{width=\linewidth}
    \caption{\justifying Effect of particle injection temperature $\Theta_{pt}$ on critical parameters, base-state fluid temperature and its stratification for H2--T2: a) variation in critical Rayleigh number ${Ra}_c$, here, the dots represent the data from \cite{prakhar2021linear}, b) variation in critical wave number $k_c$, c) variation in base-state fluid temperature, and d) variation in unstably stratified layer thickness $\delta_{st}$ with particle injection temperature for ${\Rey}_p=1$, $\delta=0.01$, $\Phi_0={10^{-4}}$, ${R}=800$, ${E}=3385$, and $\Pran=0.71$ for all graphs.}
    \label{fig:tps vs rac and kc h2t2}
\end{figure}

The effect of particle injection temperature $\Theta_{pt}$ on the critical parameters and the base-state fluid temperature is shown in figure \ref{fig:tps vs rac and kc h2t2}(a-d). In figure \ref{fig:tps vs rac h2t2}, the present data is compared with the existing work by \cite{prakhar2021linear}. It is observed that with an increase in $\Theta_{pt}$, the critical Rayleigh number ${Ra}_c$ increases in the initial part of figure \ref{fig:tps vs rac h2t2}. The explanation for this is that the particles act as the sources of heat, and by increasing their temperature, they tend to reduce the thermal stratification inside the domain, leading to increasing the stability of the flow. However, from figure \ref{fig:basestate Theta0 h2t2}, it is clear that as the particle temperature increases beyond $\Theta_{pt}>1$, the strong negative base-state temperature gradients start to appear in the upper part of the domain and favour the instability. The effect of $\Theta_{pt}$ on the critical wave number is shown in figure \ref{fig:tps vs kc h2t2}. As $\Theta_{pt}$ is increased from -1 to 2, $k_c$ decreases until around $\Theta_{pt}\approx0.5$ and then increases monotonically. This can be explained by looking at the variation in unstably stratified layer thickness $\delta_{st}$ in the base-state temperature profile shown in figure \ref{fig:basestate Theta0 h2t2} and figure \ref{fig:basestate stratification thickness h2t2}. For $\Theta_{pt}=-1$ in figure \ref{fig:basestate Theta0 h2t2}, from $z\approx-0.26$ to $-0.5$ the fluid is unstably stratified over thickness $\delta_{st}$ and in the remaining domain $\Theta_{pt}$ has a symmetric distribution. A similar distribution exists for $\Theta_{pt}=2$, but the unstably stratified layer exists near the cold top surface. The thickness of this unstably stratified layer increases $\Theta_{pt}=-1$ to $\Theta_{pt}\approx0$ and maintains a constant total height $\delta_{st}=1$ from $\Theta_{pt}\approx0$ to $\Theta_{pt}\approx1.25$ and subsequently increases monotonically for $\Theta_{pt}\approx1.25$ to $\Theta_{pt}=2$. It should be noted that the non-dimensional temperature difference across all these unstably stratified layers is maintained at 1. Hence, an increase in $\delta_{st}$ value with the same temperature difference results in a lower temperature gradient favouring stability. This explains the initial increase in ${Ra}_c$ value in figure \ref{fig:tps vs rac h2t2}. From $\Theta_{pt}\approx1.25$ to $\Theta_{pt}=2$, the unstable stratified layer thickness reduces, maintaining the same temperature difference across its length, which increases the negative temperature gradient and favours the instability. Hence, ${Ra}_c$ reduces from $\Theta_{pt}\approx1.25$ on-wards. It should be noted that $k_c$ represents the length scale for the onset of convection, and it changes with particle volume fraction $\Phi_0$ and size $\delta$. However, in this case, we keep $\Phi_0$ and $\delta$ constants and vary only $\Theta_{pt}$. Hence, the explanation for the non-monotonic variation of $k_c$ with $\Theta_{pt}$ can be deduced from the variation of $\delta_{st}$ with $\Theta_{pt}$ shown in figure \ref{fig:basestate stratification thickness h2t2}. As $\delta_{st}$ increases, the length scale at the onset of convection increases, leading to a decrease in $k_c$ and vice-versa.

\section{Formulation of finite amplitude equations}\label{sec:amplitude equation formulation}
In the present work, the finite amplitude expansions are given based on the analysis of \cite{stuart1960non,yao1992finite,khandelwal2015weakly,sharma2018finite,aleria2024finite}. The Fourier expansion of fluid temperature $\theta$ in separable form is
\begin{equation}\label{eq:perturbation series}
    \begin{split}
        \theta(x,\,z,\,t)&=\Theta(z,\,\tau)\mathbb{E}^0+\hat{\theta}_1\left(z,\,\tau\right)\mathbb{E}^1+\hat{\theta}_2\left(z,\,\tau\right)\mathbb{E}^2+\dots+\text{c.\,c.},\\
        &=\mathbb{E}^0\left[\Theta_0(z)+c_i\abs{B(\tau)}^2\Theta_1(z)+\order{c_i^2}\right]\\
        &\qquad+\mathbb{E}^1\left[c_i^{1/2}B(\tau)\theta_{10}(z)+c_i^{3/2}B(\tau)\abs{B(\tau)}^2\theta_{11}(z)+\order{c_i^{5/2}}\right]\\
        &\qquad\qquad+\mathbb{E}^2\left[c_iB(\tau)^2\theta_{20}(z)+\order{c_i^2}\right]+\dots+\text{c.\,c,}
    \end{split}
\end{equation}
where $\mathbb{E}^j=\exp{\displaystyle j\left[\mbox{i}k_c\left(x-c_rt\right)\right]}$, for $j=\left\{0,\,1,\,2\right\}$; $k_c$ is the wavenumber corresponding to the critical Rayleigh number ${Ra}_c$, and $c_r$ is the real part of $c$ corresponding to the most unstable disturbance. The amplitude function $B$ will be obtained from the Landau equation, and c.c. stands for the complex conjugate of the complex-valued functions (the second and the third term). 

The evolution equation for the slowly varying amplitude $B$ is derived using the method of multiple time scales. Hence, two distinct time scales (a fast time scale $(t)$ and a slow time scale $(\tau)$) are used in the nonlinear stability analysis of the flow. As in linear stability analysis, the exponential variation of disturbance amplitude is associated with the fast time scale. As the disturbance amplitude grows exponentially and attains a finite amplitude, the nonlinear terms become important. This leads to deviation of temporal exponential growth of the disturbances. Hence, another time scale (the slow time scale) is needed to capture the effect of nonlinearities of different orders. The slow time scale is defined as $\tau=c_it$, with $c_i$ being the growth rate of the most unstable disturbance obtained from the linear stability analysis. Hence, the derivative with respect to $t$ of a function $f(t)=F(t,\,\tau)$ is given by
\begin{equation}\label{eq:time derivative}
    \pdv{f}{t}=\pdv{F}{t}+c_i\pdv{F}{\tau}.
\end{equation}
Thus, for a function of the form $F(t,\, \tau)$, both $t$ and $\tau$ are treated as if there are independent variables. The justification of equation (\ref{eq:perturbation series}) is as follows: from the linear theory, it is known that the growth rate $c_i$ is linearly proportional to $\left({Ra}-{Ra}_c\right)/{Ra}_c$, the Rayleigh number difference from the neutral curve. Using a finite-amplitude stability analysis for the Rayleigh-B\'enard convection, \cite{schluter1965stability} and for the non-isothermal flow in a vertical annulus, \cite{yao1992finite} showed that the square of equilibrium amplitude $A_e^2$ is proportional to $\left({Ra}-{Ra}_c\right)/{Ra}_c$ and consequently the equilibrium amplitude is proportional to ${c_i^{1/2}}$ with a proportionality constant which is a weak function of ${Ra}$ near ${Ra}_c$. Hence, the amplitude of the disturbance fluid temperature ${\theta}_{10}$ with wave number $k_c$ is the order of $c_i^{1/2}$. The amplitude of the disturbances of higher harmonics is chosen based on the number of interactions between lower-order harmonics. For example, the harmonic with $2k_c$ wave number needs two disturbances with wave number $k_c$, which gives its amplitude of order $c_i$. Similarly, the higher-order correction to the base state results from the Reynolds stress due to the interaction of $\mathbb{E}^1$ and its complex conjugate $\mathbb{E}^{-1}$ with an amplitude of order $c_i$ \citep{stewartson1971non}. 

\subsection{Derivation of cubic Landau equation}\label{sec:Landau equation derivation}
Substitute the perturbation series similar to the equation (\ref{eq:perturbation series}) for all the dependent variables in equation (\ref{eq:fluid mass balance})--(\ref{eq:particle energy balance}) and separate the harmonic components, the equations for the harmonic $\mathbb{E}^0$ are given as follows:
\begin{eqnarray}\label{eq:E0 uz equation h2t2}
        &&\quad\mathcal{L}_z\qty(P_0,\,\Theta_0,\,v_0,\,\Phi_0,\,{R},\,{St}_m)
        +c_i\abs{B}^2\qty{\mathcal{L}_z\qty(P_1,\,\Theta_1,\,-V_1,\,\Phi_0,\,{R},\,{St}_m)+\frac{{R}v_0}{{St}_m}\Phi_1}\nonumber\\
        &&\qquad +c_i\abs{B}^2\qty{2\dv{z}\qty(\Tilde{u}_{z10}u_{z10})+\frac{{R}}{{St}_m}\qty{\phi_{10}\qty(\Tilde{u}_{z10}-\Tilde{v}_{z10})+\Tilde{\phi}_{10}\qty(u_{z10}-v_{z10})}}=\order{c_i^2},\\ \label{eq:E0 theta equation h2t2}
        &&\quad\mathcal{L}_\theta\qty(\Theta_0,\,\Theta_{p0},\,\Phi_0,\,{Ra},\,{\Pran},\,{E},\,{St}_{th})\nonumber\\
        &&\qquad+c_i\abs{B}^2\qty{\mathcal{L}_\theta\qty(\Theta_1,\,\Theta_{p1},\,\Phi_0,\,{Ra},\,{\Pran},\,{E},\,{St}_{th})-\frac{{E}\qty(\Theta_0-\Theta_{p0})}{{St}_{th}}\Phi_1}\nonumber\\
        &&\qquad\qquad-c_i\abs{B}^2\left\{\dv{z}\qty(\Tilde{\theta}_{10}u_{z10}+\theta_{10}\Tilde{u}_{z10})\right.\nonumber\\
        &&\qquad\qquad\qquad\left.+\frac{{E}}{{St}_{th}}\qty(\phi_{10}\qty(\Tilde{\theta}_{10}-\Tilde{\theta}_{p10}) +\Tilde{\phi}_{10}\qty(\theta_{10}-\theta_{p10}))\right\}=\order{c_i^2},\\ \label{eq:E0 phi equation h2t2}
        &&\quad c_i\abs{B}^2\qty{\Phi_0\dv{V_1}{z}-v_0\dv{\Phi_1}{z}+\dv{z}\qty(\phi_{10}\Tilde{v}_{z10})+\dv{z}\qty(\Tilde{\phi}_{10}v_{z10})}=\order{c_i^2},\\
        &&\quad c_i\abs{B}^2\qty{\qty(v_0\dv{V_1}{z}-\frac{V_1}{{St}_m})+\text{i}k_c\qty(v_{x10}\Tilde{v}_{z10}-\Tilde{v}_{x10}v_{z10})-\dv{z}\qty(v_{z10}\Tilde{v}_{z10})}=\order{c_i^2},\\ \label{eq:E0 thetap equation h2t2}
        &&\quad\mathcal{L}_{p\theta}\qty(\Theta_0,\,\Theta_{p0},\,v_0,\,{St}_{th})+c_i\abs{B}^2\qty{\mathcal{L}_{p\theta}\qty(\Theta_1,\,\Theta_{p1},\,v_0,\,{St}_{th})-V_1\dv{\Theta_{p0}}{z}}\nonumber\\
        &&\qquad -c_i\abs{B}^2\qty{v_{z10}\dv{\Tilde{\theta}_{p10}}{z}+\Tilde{v}_{z10}\dv{\theta_{p10}}{z}+\text{i}k_c\qty(\Tilde{v}_{x10}\theta_{p10}-v_{x10}\Tilde{\theta}_{p10})}=\order{c_i^2}
\end{eqnarray}
for $z\in\qty(-1/2,\,1/2)$. Where the operators $\mathcal{L}_z$, $\mathcal{L}_\theta$, and $\mathcal{L}_{p\theta}$ are given by equations (\ref{eq:basestate uz equation}), (\ref{eq:basestate theta equation}), and (\ref{eq:basestate thetap equation}), respectively. The boundary conditions for $\Theta_1$, $V_1$, $\Theta_{p1}$, and $\Phi_1$ are given by
\begin{align}\label{eq:bc for distortion h2t2}
    \text{At }z&=-1/2:\text{ }\Theta_1=0,\\
    \text{At }z&=1/2:\text{ }\Theta_1=V_1=\Theta_{p1}=\Phi_1=0.
\end{align}
Similarly, the equations for the harmonic $\mathbb{E}^1$ are given by
\begin{eqnarray}\label{eq:E1 continuity equation}
    &&\quad\qty(c_i)^{1/2}B\mathscr{L}_0\qty(k_c,\,u_{x10},\,u_{z10})+\qty(c_i)^{3/2}B\abs{B}^2\mathscr{L}_0\qty(k_c,\,u_{x11},\,u_{z11})=\order{c_i^{5/2}},\\
    &&\quad\qty(c_i)^{1/2}B\mathscr{L}_x\qty(k_c,\,c,\,u_{x10},\,p_{10},\,v_{x10},\,\Phi_0,\,{Ra},\,{\Pran},\,{R},\,{St}_m)\nonumber\\
    &&\qquad+ \qty(c_i)^{3/2}B\abs{B}^2\mathscr{L}_x\qty(k_c,\,c,\,u_{x11},\,p_{11},\,v_{x11},\,\Phi_0,\,{Ra},\,{\Pran},\,{R},\,{St}_m)\nonumber\\
    &&\qquad\qquad -\qty(c_i)^{3/2}\qty(B\abs{B}^2\mathcal{G}_x+\qty{\dv{B}{\tau}-k_cB}u_{x10})=\order{c_i^{5/2}},\\
    &&\quad\qty(c_i)^{1/2}B\mathscr{L}_z\qty(k_c,\,c,\,u_{z10},\,p_{10},\,\theta_{10},\,v_{z10},\,\phi_{10},\,v_0,\,\Phi_0,\,{Ra},\,{\Pran},\,{R},\,{St}_m)\nonumber\\
    &&\qquad\qquad+ \qty(c_i)^{3/2}B\abs{B}^2\mathscr{L}_z\qty(k_c,\,c,\,u_{z11},\,p_{11},\,\theta_{11},\,v_{z11},\,\phi_{11},\,v_0,\,\Phi_0,\,{Ra},\,{\Pran},\,{R},\,{St}_m)\nonumber\\
    &&\qquad\qquad\qquad -\qty(c_i)^{3/2}\qty(B\abs{B}^2\mathcal{G}_z+\qty{\dv{B}{\tau}-k_cB}u_{z10})=\order{c_i^{5/2}},\\
    &&\quad\qty(c_i)^{1/2}B\mathscr{L}_\theta\qty(k_c,\,c,\,u_{z10},\,\theta_{10},\,\theta_{p10},\,\phi_{10},\,\Theta_{0},\,\Theta_{p0},\,\Phi_0,\,{Ra},\,{\Pran},\,{E},\,{St}_{th})\nonumber\\
    &&\qquad\qquad+\qty(c_i)^{3/2}B\abs{B}^2\mathscr{L}_\theta\qty(k_c,\,c,\,u_{z11},\,\theta_{11},\,\theta_{p11},\,\phi_{11},\,\Theta_0,\,\Theta_{p0},\,\Phi_0,\,{Ra},\,{\Pran},\,{E},\,{St}_{th})\nonumber\\
    &&\qquad\qquad\qquad -\qty(c_i)^{3/2}\qty(B\abs{B}^2\mathcal{G}_\theta+\qty{\dv{B}{\tau}-k_cB}\theta_{10})=\order{c_i^{5/2}},\\
    &&\quad\qty(c_i)^{1/2}B\mathscr{L}_{\phi}\qty(k_c,\,c,\,v_{x10},\,v_{z10},\,\phi_{10},\,v_0,\,\Phi_0)\nonumber\\
    &&\qquad\qquad +\qty(c_i)^{3/2}B\abs{B}^2\mathscr{L}_{\phi}\qty(k_c,\,c,\,v_{x11},\,v_{z11},\,\phi_{11},\,v_0,\,\Phi_0)\nonumber\\
    &&\qquad\qquad\qquad-\qty(c_i)^{3/2}\qty(B\abs{B}^2\mathcal{G}_\phi+\qty{\dv{B}{\tau}-k_cB}\phi_{10})=\order{c_i^{5/2}},\\
    &&\quad\qty(c_i)^{1/2}\mathscr{L}_{px}\qty(k_c,\,c,\,u_{x10},\,v_{x10},\,v_0,\,{St}_m)+\qty(c_i)^{3/2}B\abs{B}^2\mathscr{L}_{px}\qty(k_c,\,c,\,u_{x11},\,v_{x11},\,v_0,\,{St}_m)\nonumber\\
    &&\qquad -\qty(c_i)^{3/2}\qty(B\abs{B}^2\mathcal{G}_{px}+\qty{\dv{B}{\tau}-k_cB}v_{x10})=\order{c_i^{5/2}},\\
    &&\quad\qty(c_i)^{1/2}B\mathscr{L}_{pz}\qty(k_c,\,c,\,u_{z10},\,v_{z10},\,v_0,\,{St}_m)+\qty(c_i)^{3/2}B\abs{B}^2\mathscr{L}_{pz}\qty(k_c,\,c,\,u_{z11},\,v_{z11},\,v_0,\,{St}_m)\nonumber\\
    &&\qquad -\qty(c_i)^{3/2}\qty(B\abs{B}^2\mathcal{G}_{pz}+\qty{\dv{B}{\tau}-k_cB}v_{z10})=\order{c_i^{5/2}},\\ \label{eq:E1 thetap equation}
    &&\quad\qty(c_i)^{1/2}B\mathscr{L}_{p\theta}\qty(k_c,\,c,\,\theta_{10},\,v_{z10},\,\theta_{p10},\,v_0,\,\Theta_{p0},\,{St}_{th})\nonumber\\
    &&\qquad\qquad+\qty(c_i)^{3/2}B\abs{B}^2\mathscr{L}_{p\theta}\qty(k_c,\,c,\,\theta_{11},\,v_{z11},\,\theta_{p11},\,v_0,\,\Theta_{p0},\,{St}_{th})\nonumber\\
    &&\qquad\qquad\qquad -\qty(c_i)^{3/2}\qty(B\abs{B}^2\mathcal{G}_{p\theta}+\qty{\dv{B}{\tau}-k_cB}\theta_{p10})=\order{c_i^{5/2}}
\end{eqnarray}
for $z\in\qty(-1/2,\,1/2)$ and the operators $\mathscr{L}_0$, $\mathscr{L}_x$, $\mathscr{L}_z$, $\mathscr{L}_\theta$, $\mathscr{L}_{\phi}$, $\mathscr{L}_{px}$, $\mathscr{L}_{pz}$, and $\mathscr{L}_{p\theta}$ are the linear stability operators given by the equations (\ref{eq:linear continuity equation})-(\ref{eq:thetap linear equation}) with boundary conditions given by the equations (\ref{eq:linear bc at z=1/2})--(\ref{eq:linear bc at z=-1/2}) for both $\qty(u_{x10},\,u_{z10},\,\theta_{10},\,v_{x10},\,v_{z10},\,\theta_{p10},\,\phi_{10})$ and $\qty(u_{x11},\,u_{z11},\,\theta_{11},\,v_{x11},\,v_{z11},\,\theta_{p11},\,\phi_{11})$.  Where the scalar functions $\mathcal{G}_x$, $\mathcal{G}_z$, $\mathcal{G}_\theta$, $\mathcal{G}_{px}$, $\mathcal{G}_{pz}$, $\mathcal{G}_\phi$, and $\mathcal{G}_{p\theta}$ are defined in appendix \ref{appen:scalar functions}.
 
 The equations for the harmonic $\mathbb{E}^2$ are given by
 \begin{eqnarray}\label{eq:E2 continuity equation}
        &&\quad c_iB^2\mathscr{L}_0\qty(2k_c,\,u_{x20},\,u_{z20})=\order{c_i^2},\\
        &&\quad c_iB^2\mathscr{L}_x\qty(2k_c,\,u_{x20},\,p_{20},\,v_{x20},\,\Phi_0,\,{Ra},\,{\Pran},\,{R},\,{St}_m)\nonumber\\
        &&\qquad-c_iB^2\qty{u_{z10}\dv{u_{x10}}{z}-u_{x10}\dv{u_{z10}}{z}+\frac{{R}\phi_{10}}{{St}_m}\qty(u_{x10}-v_{x10})}=\order{c_i^2},\\
        &&\quad c_iB^2\qty{\mathscr{L}_z\qty(2k_c,\,c,\,u_{z20},\,p_{20},\,\theta_{20},\,v_{z20},\,\Phi_0,\,{Ra},\,{\Pran},\,{R},\,{St}_m)-\frac{{R}\phi_{10}}{{St}_m}\qty(u_{z10}-u_{x10})}\nonumber\\
        &&\qquad=\order{c_i^2},\\
        &&\quad c_iB^2\mathscr{L}_{\theta}\qty(2k_c,\,c,\,u_{z20},\,\theta_{20},\,\theta_{p20},\,\phi_{20},\,\Theta_0,\,\Theta_{p0},\,\Phi_0,\,{Ra},\,{\Pran},\,{E},\,{St}_{th})\nonumber\\
        &&\qquad-c_iB^2\qty{u_{z10}\dv{\theta_{10}}{z}-\theta_{10}\dv{u_{z10}}{z}+\frac{{E}\phi_{10}}{{St}_{th}}\qty(\theta_{10}-\theta_{p10})}=\order{c_i^2},\\
        &&\quad c_iB^2\mathscr{L}_{\phi}\qty(2k_c,\,c,\,v_{x20},\,v_{z20},\,\phi_{20},\,v_0,\,\Phi_0)
        -c_iB^2\qty{2\text{i}k_c\phi_{10}v_{x10}+\dv{z}\qty(\phi_{10}v_{z10})}\nonumber\\
        &&\qquad=\order{c_i^2},\\
        &&\quad c_iB^2\qty{\mathscr{L}_{px}\qty(2k_c,\,c,\,u_{x20},\,v_{x20},\,v_0,\,{St}_m)-\qty(\text{i}k_cv_{x10}^2+v_{z10}\dv{v_{x10}}{z})}=\order{c_i^2},\\
        &&\quad c_iB^2\qty{\mathscr{L}_{pz}\qty(2k_c,\,c,\,u_{z20},\,v_{z20},\,v_0,\,{St}_m)-\qty(\text{i}k_cv_{x10}v_{z10}+v_{z10}\dv{v_{z10}}{z})}=\order{c_i^2},\\ \label{eq:E2 thetap equation}
        &&\quad c_iB^2\qty{\mathscr{L}_{p\theta}\qty(2k_c,\,c,\,\theta_{20},\,v_{z20},\,\theta_{p20},\,v_0,\,\Theta_{p0},\,{St}_{th})-\qty(\mbox{i}k_cv_{x10}\theta_{p10}+v_{z10}\dv{\theta_{p10}}{z})}\nonumber\\
        &&\qquad=\order{c_i^2}
 \end{eqnarray}
for $z\in(-1/2,\,1/2)$ and the boundary conditions are similar to the (\ref{eq:linear bc at z=1/2})--(\ref{eq:linear bc at z=-1/2}). Here $\sim$ represents the complex conjugate. The first Landau coefficient does not depend on the higher-order harmonics ($\mathbb{E}^3,\,\mathbb{E}^4,\,$ etc.). Therefore, those terms are ignored in the series (\ref{eq:perturbation series}).

The system of equations at different harmonics (\ref{eq:E0 uz equation h2t2})--(\ref{eq:E0 thetap equation h2t2}), (\ref{eq:E1 continuity equation})--(\ref{eq:E1 thetap equation}), and (\ref{eq:E2 continuity equation})--(\ref{eq:E2 thetap equation}) are solved at increase orders of $c_i$. At $\order{c_i^0}$, $\mathbb{E}^0$ harmonic equations are identical to the basic-flow equations (\ref{eq:basestate uz equation})--(\ref{eq:basestate thetap equation}). At $\order{c_i}^{1/2}$, $\mathbb{E}^1$ equations are similar to that of linear stability equations (\ref{eq:linear continuity equation})--(\ref{eq:thetap linear equation}) and the $\mathbb{E}^0$ and $\mathbb{E}^2$ do not contribute at this order. The functions $u_{x10}$, $u_{z10}$, $\theta_{10}$, $v_{x10}$, $v_{z10}$, $\theta_{p10}$, and $\phi_{10}$ are the eigenfunctions of linear stability equations at a particular wave number and Rayleigh number $Ra$. At $\order{c_i}$, $\mathbb{E}^0$ and $\mathbb{E}^2$ generate the system of non-homogeneous equations for basic-flow distortion functions $\Theta_1$, $\Theta_{p1}$, $V_1$, and $\Phi_1$ and the functions $u_{x20}$, $u_{z20}$, $\theta_{20}$, $v_{x20}$, $v_{z20}$, $\theta_{p20}$, and $\phi_{20}$, respectively. These equations contain the non-homogeneous part made up of known functions $u_{x10}$, $u_{z10}$, $\theta_{10}$, $v_{x10}$, $v_{z10}$, $\theta_{p10}$, $\phi_{10}$ and their derivatives which are obtained at lower order computations. At $\order{c_i^{3/2}}$, $\mathbb{E}^1$ harmonic results in a non-homogeneous system of equations with linear stability operators acting on the functions $u_{x11}$, $u_{z11}$, $\theta_{11}$, $v_{x11}$, $v_{z11}$, $\theta_{p11}$, and $\phi_{11}$. However, the right-hand side non-homogeneous part contains the terms proportional to the unknown terms $dB/d\tau$, $B$, and $B\abs{B}^2$ with coefficients depending on functions known from the lower order analysis. Fredholm alternative solvability condition is imposed on the non-homogenous right-hand side of $\mathbb{E}^1$ harmonic at $\order{c_i^{3/2}}$ to obtain a non-trivial solution. To impose the solvability condition, the solution to the homogeneous adjoint system (see appendix \ref{appen:linear adjoint equations}) corresponding to the linear stability operator is required. Accordingly, the non-homogeneous right-hand side must be orthogonal to the adjoint functions $\hat{u}_x^\dag$,  $\hat{u}_z^\dag$,  $\hat{\theta}^\dag$,  $\hat{v}_x^\dag$,  $\hat{v}_z^\dag$,  $\hat{\theta}_{p}^\dag$, and  $\hat{\phi}^\dag$. The orthogonality is imposed by multiplying the right-hand side of the system of linear equations of $\mathbb{E}^1$ harmonic functions at $\order{c_i^{3/2}}$ with $\displaystyle\qty(\hat{p}^\dag,\, \hat{u}_{x}^\dag,\,\hat{u}_z^\dag,\,\hat{\theta}^\dag, \,\hat{\phi}^\dag,\,\hat{v}_x^\dag,\hat{v}_z^\dag,\hat{\theta}_p^\dag)$, and integrate with respect to $z$ from $-1/2$ to $1/2$ which gives the following cubic Landau equation,
\begin{equation}\label{eq:Landau cubic eq.}
    \dv{B}{\tau}=k_cB + a_1B\abs{B}^2,
\end{equation}
where $a_1$ is the Landau constant defined as
\begin{equation}\label{eq:Landau constant}
    a_1=-\int_{-1/2}^{1/2}\qty(\mathcal{G}_x\hat{u}_{x}^\dag+\mathcal{G}_z\hat{u}_{z}^\dag+\mathcal{G}_\theta \hat{\theta}^\dag+\mathcal{G}_\phi\hat{\phi}^\dag+\mathcal{G}_{px}\hat{v}_{x}^\dag+\mathcal{G}_{pz}\hat{v}_{z}^\dag+\mathcal{G}_{p\theta} \hat{\theta}_p^\dag)\,dz,
\end{equation}
which represents the first correction to the growth rate given by linear stability analysis.

By changing the variables as $A=c_i^{1/2}B$ and $\tau=c_it$ in (\ref{eq:Landau cubic eq.}), 
\begin{equation}\label{eq:dA/dt}
    \dv{A}{t}=k_cc_iA+a_1A^2\Tilde{A},
\end{equation}
and its corresponding complex conjugate equation is
\begin{equation}\label{eq:dtilde(A)/dt}
    \dv{\Tilde{A}}{t}=k_cc_i\Tilde{A}+\Tilde{a}_1\Tilde{A}^2A,
\end{equation}
where $k_cc_i$ is the growth rate from the linear stability analysis, and $A$ is the physical amplitude of the wave. Multiply (\ref{eq:dA/dt}) with $\Tilde{A}$ and (\ref{eq:dtilde(A)/dt}) with $A$ and add them,
\begin{equation}\label{eq:abs(A) eq.}
    \dv{\abs{A}^2}{t}=2k_cc_i\abs{A}^2+2\Re{a_1}\abs{A}^4,
\end{equation}
where $\Re{a_1}$ is the real part of $a_1$. The equation (\ref{eq:abs(A) eq.}) has an equilibrium amplitude, $A_e$ as a solution, if
\begin{equation}
    \dv{\abs{A}}{t}=0.
\end{equation}
Consequently, three possible equilibrium amplitudes exist,

\refstepcounter{equation}
$$
A_e=0\qand A_e=\pm\sqrt{-k_cc_i/\Re{a_1}}.
\eqno{(\theequation{\mathit{a},\,\mathit{b}})}\label{eq:equilibrium amplitude}
$$
Where $A_e=0$ represents the steady base flow, which is stable for ${Ra}<{Ra}_c$ and unstable for ${Ra}>{Ra}_c$, here, ${Ra}_c$ is the critical Rayleigh number obtained from linear stability analysis. From the equation (\ref{eq:equilibrium amplitude}b), the existence of a finite amplitude solution is guaranteed if $k_cc_i$ and $\Re{a_1}$ have opposite signs \citep{drazin2004hydrodynamic, shukla2011weakly}. Therefore, there are two possibilities: first, the growth rate is positive(for ${Ra}>{Ra}_c$), and the real part of the Landau constant is negative; second, the growth rate is negative (for ${Ra}<{Ra}_c$), and the real part of the Landau constant is positive. The first possibility leads to a supercritical pitchfork bifurcation, whereas the second possibility leads to a subcritical pitchfork bifurcation.

Using equilibrium amplitude definition (\ref{eq:equilibrium amplitude}b), (\ref{eq:abs(A) eq.}) can be rewritten as
\begin{equation}
    \dv{\abs{A}^2}{t}=2\Re{a_1}\abs{A}^2\qty(\abs{A}^2-A_e^2)
\end{equation}
where $A_e^2=-\qty(k_cc_i)/\Re{a_1}$.
The solution for the above equation is
\begin{equation}\label{eq:mod(A)^2 solution}
    \abs{A}^2=\frac{A_e^2}{1+\displaystyle\qty(\frac{A_e^2}{\abs{A_0}^2}-1)e^{\displaystyle-2k_cc_it}},
\end{equation}
where $\abs{A_0}$ is the value of $\abs{A}$ at $t=0$. From the equation (\ref{eq:mod(A)^2 solution}), it is clear that when ${Ra}>{Ra}_c$, as $t\to\infty$, $\abs{A}\to A_e$. Hence, irrespective of the initial amplitude $\abs{A_0}$, the amplitude eventually tends to the equilibrium amplitude $A_e$. 
Multiply (\ref{eq:dA/dt}) with $1/\Tilde{A}$ and (\ref{eq:dtilde(A)/dt}) with $-A/\Tilde{A}^2$ and add them
\begin{equation}
    \dv{t}\qty(\frac{A}{\Tilde{A}})=2\text{i}A^2\Im{a_1}
\end{equation}
where $\Im{a_1}$ is the imaginary part of $a_1$. The above equation can be rewritten as the following
\begin{equation}
    \frac{\text{d}\displaystyle\qty(\frac{A}{\Tilde{A}})}{\displaystyle\qty(\frac{A}{\Tilde{A}})}=2{i}\Im{a_1}\abs{A}^2\,dt.
\end{equation}
Hence, the solution for $A$ is given by
\begin{equation}\label{eq:A(t) in phase angle form}
    A=\displaystyle\abs{A}\exp{\displaystyle\text{i}\Im{a_1}\int_0^t\abs{A}^2\,dt},
\end{equation}
where $\abs{A}\qty(t)$ is given by the square root of (\ref{eq:mod(A)^2 solution}).
The closed-form solution for amplitude function $A\qty(t)$ is obtained by integrating the above equation,
\begin{equation}\label{eq:expression of A(t)}
    \displaystyle \frac{A(t)}{A_0}=\frac{\abs{A}}{\abs{A_0}}\qty{\frac{\abs{A}}{\abs{A_0}}e^{\displaystyle-k_cc_it}}^{\displaystyle\frac{\text{i}\Im{a_1}}{\Re{a_1}}}
\end{equation}
where $\abs{A}(t)$ is obtained from the equation (\ref{eq:mod(A)^2 solution}).

\subsection{Heat flux balance}\label{sec:heat flux balance}
The solution to the Landau equation given by (\ref{eq:mod(A)^2 solution}) and (\ref{eq:expression of A(t)}) shows the existence of steady-state solution as $t\to\infty$. Hence, in this section, we derive an equation for the average heat flux balance at steady-state. At steady-state, integrating the fluid energy equation (\ref{eq:fluid energy equation}) in the domain $\qty(x,\,z)\in\qty{\qty(-0.5,\,0.5)\times\qty(-\pi/k_c,\,\pi/k_c)}$ yields,
\begin{equation}\label{eq:avg. fluid energy}
    \begin{split}
        \int_{z=-0.5}^{0.5}\int_{x=-\pi/k_c}^{\pi/k_c}\qty(u_x\pdv{\theta}{x}+u_z\pdv{\theta}{z})\,dx\,dz\\
        \qquad=\int_{z=-0.5}^{0.5}\int_{x=-\pi/k_c}^{\pi/k_c}\qty(\frac{1}{\sqrt{{RaPr}}}\qty(\pdv[2]{\theta}{x}+\pdv[2]{\theta}{z})-\frac{{E}\phi}{{St}_{th}}\qty(\theta_p-\theta_p))\,dx\,dz.
    \end{split}
\end{equation}

Since $u_x$, $u_z$, $\theta$ and $\theta_p$ are periodic functions in $x$ with period $2\pi/k$, the non-zero contribution comes only from the $\mathbb{E}^0$ harmonic. Hence, the above equation (\ref{eq:avg. fluid energy}) is essentially equivalent to the integration of $\mathbb{E}^0$ harmonic (\ref{eq:E0 theta equation h2t2}) in $z\in\qty(-0.5,\,0.5)$. Therefore, after substituting $A_e^2=c_i\abs{B}^2$, $\Theta=\Theta_0+A_e^2\Theta_1$, and $\Theta_p=\Theta_{p0}+A_e^2\Theta_{p1}$ in (\ref{eq:E0 theta equation h2t2}),

\begin{align}\label{eq:integral energy balance h2t2 intermediate}
	\quad\int_{z=-0.5}^{0.5}\dv[2]{\Theta}{z}\,dz&={E}\sqrt{{Ra\Pran}}\int_{z=-0.5}^{0.5}\left\{\Phi_0\frac{\Theta-\Theta_p}{{St}_{th}}+A_e^2\Phi_1\frac{\Theta_0-\Theta_{p0}}{{St}_{th}}\right.\nonumber\\
	&\qquad\left.+A_e^2\dv{z}\qty(\Tilde{\theta}_{10}u_{z10}+\theta_{10}\Tilde{u}_{z10})\right\}\,dz,
\end{align}
where the integral of the last term on the right-hand side goes to zero because $u_{z10}=\theta_{10}=0$ at $z=-0.5$ and 0.5. From (\ref{eq:E0 thetap equation h2t2}) we have the following identities, 
\begin{eqnarray}
    \frac{\Theta-\Theta_p}{{St}_{th}}&=&-v_0\dv{\Theta_p}{z}+A_e^2V_1\dv{\Theta_{p0}}{z},\\
    \frac{\Theta_0-\Theta_{p0}}{{St}_{th}}&=&-v_0\dv{\Theta_{p0}}{z}.
\end{eqnarray}
Hence, the equation (\ref{eq:integral energy balance h2t2 intermediate}) simplified to
\begin{equation}\label{eq:heat flux balance 1}
    \llbracket{Nu}\rrbracket=v_0{E}\Phi_0\sqrt{{Ra\Pran}}\displaystyle\llbracket \Theta_p \rrbracket+A_e^2\int_{z=-0.5}^{0.5}\qty(\Phi_0V_1-\Phi_1v_0)\dv{\Theta_{p0}}{z}\,dz,
\end{equation}
where the bracket is defined for the function $f(z)$ as $\llbracket f \rrbracket=f(0.5)-f(-0.5)$ and ${Nu}_h={Nu}(-0.5)=-\eval{\dv{\Theta}{z}}_{z=-0.5}$ and ${Nu}_c={Nu}(0.5)=-\eval{\dv{\Theta}{z}}_{z=0.5}$ are the Nusselt numbers (or non-dimensional heat fluxes) at the hot and cold surfaces, respectively. It can be shown that the solution to the linearized volume fraction equation is always zero (see appendix \ref{appen:linear particle volume fraction solution}). Hence, from the equation (\ref{eq:E0 phi equation h2t2}),
\begin{equation}\label{eq:relation between the Phi and V}
    \dv{z}\qty(\Phi_0V_1-\Phi_1v_0)=0
\end{equation}
for $z\in\qty(-0.5,\,0.5)$. Therefore, from the (\ref{eq:heat flux balance 1}) and (\ref{eq:relation between the Phi and V}), the net flux balance is given by
\begin{equation}
     \llbracket{Nu}\rrbracket=v_0{E}\Phi_0\sqrt{{Ra\Pran}}\displaystyle\llbracket \Theta_p \rrbracket,
\end{equation}
where the proportionality constant $v_0{E}\Phi_0\sqrt{{Ra\Pran}}$ is equal ${L}^2/l$. Here, ${L}$ and $l$ are the non-dimensional length scales given by (\ref{eq:L definition}) and (\ref{eq:l definition}), respectively. Thus, the alternate form of the above equation is given by
\begin{equation}\label{eq:net thermal energy flux balance}
    \llbracket{Nu}\rrbracket=\displaystyle\llbracket Q_p^{\prime\prime} \rrbracket,
\end{equation}
where $Q_{ph}^{\prime\prime}=Q_p^{\prime\prime}(-0.5)=(l/{L}^2)\Theta_p(-0.5)$ and  $Q_{pc}^{\prime\prime}=Q_p^{\prime\prime}(0.5)=(l/{L}^2)\Theta_p(0.5)$ are the non-dimensional particle sensible heat fluxes at hot and cold surfaces, respectively. The physical significance of (\ref{eq:net thermal energy flux balance}) is that at steady-state, the net heat flux released from the hot and cold surface is equal to the net sensible heat exchanged by the particles. Therefore, for the single-phase and particle-laden flows without thermal energy two-way coupling, the heat flux at the hot surface is exactly the heat flux at the hot surface. However, for the problems with thermal energy two-way coupling, the two phases exchange the heat by following (\ref{eq:net thermal energy flux balance}) and heat fluxes $(Nu)$ at hot and cold need not be equal.

\section{Results and discussion}\label{sec:results and discussion}
We carry out the analysis of the effect of the nonlinear interactions of different harmonics on the equilibrium amplitude, rate of heat transfer and the pattern of secondary flow. All the analysis is done near the bifurcation point such that the perturbation series (\ref{eq:perturbation series}) is valid. We defined new parameters called reduced Rayleigh number $\delta_{Ra}=\qty(Ra-Ra_c)/Ra_c$ quantifies the deviation from the bifurcation point $Ra_c$.

\subsection{Effect of particle volume fraction}
\begin{table}
\centering
\begin{tabular}{ccccc}
\toprule
$\Phi_0$                     & $N$  & $c_i$ & $\Re{a_1}$ & $A_e$ \\
\midrule
\multirow{4}{*}{$10^{-5}$} & 30 & 0.0095822260252546 & -2.5036981432421341  & 0.1105637883706295     \\
                             & 40 & 0.0095822260247759 & -2.5036981431506380  & 0.1105637883698877     \\
                             & 50 & 0.0095822260222355 & -2.5036981431784873  & 0.1105637883546169     \\
                             & 60 & 0.0095822260252879 & -2.5036981430774201  & 0.1105637883744580     \\
\midrule
\multirow{4}{*}{$10^{-4}$} & 30 & 0.0080109706070805  & -3.1611749067227288  & 0.0977402064516451    \\
                             & 40 & 0.0080109706065269  & -3.1611749067584540  & 0.0977402064477159    \\
                             & 50 & 0.0080109706064721  & -3.1611749068802188  & 0.0977402064454993    \\
                             & 60 & 0.0080109706072748  & -3.1611749065909187  & 0.0977402064548681    \\ 
\midrule
\multirow{4}{*}{$10^{-3}$} & 30 & 0.0044982008778323  & -5.5363083852835402  & 0.0802866894011331    \\
                             & 40 & 0.0044982008776943  & -5.5363086210155394  & 0.0802866876906275    \\
                             & 50 & 0.0044982008777145  & -5.5363086208461425  & 0.0802866876920357    \\
                             & 60 & 0.0044982008776811  & -5.5363086210110506  & 0.0802866876905422    \\ 
\bottomrule   
\end{tabular}
\captionsetup{width=\linewidth}
\caption{\justifying Grid independence test on growth rate $(c_i)$, real part of Landau constant $\left(\Re{a_1}\right)$ and equilibrium amplitude $(A_e)$ for $\Rey_p=1$, $\delta=0.01$, ${R}=800$, ${E}=3385$, $\Pran=0.71$, $\Theta_{pt}=0$ and $\Phi_0=\qty{10^{-5},\, 10^{-4},\, 10^{-3}}$ at $\delta_{Ra}=0.1$.}
\label{tab:grid independence test for nonlin-H2T2}
\end{table}

\begin{figure}
    \centering
    \begin{subfigure}[b]{0.49\textwidth}
        \includegraphics[width=\textwidth]{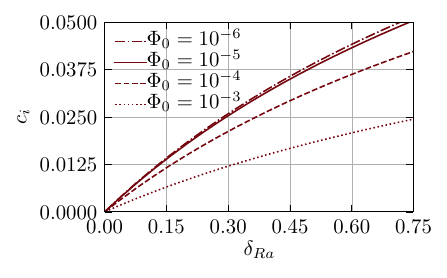}
        \caption{}
        \label{fig:ci vs ra h2t2} 
    \end{subfigure}
    \hfill
    \begin{subfigure}[b]{0.49\textwidth}
        \includegraphics[width=\textwidth]{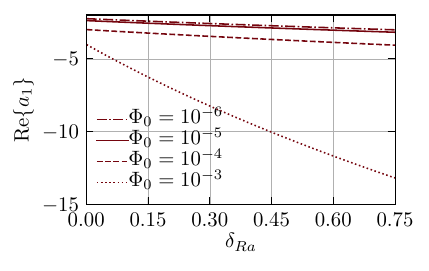}
        \caption{}
        \label{fig:a1r vs ra h2t2}
    \end{subfigure}
    \hfill
    \begin{subfigure}[b]{0.49\textwidth}
        \includegraphics[width=\textwidth]{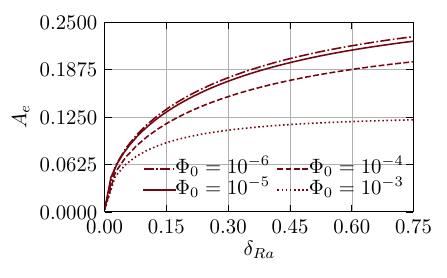}
        \caption{}
        \label{fig:Ae vs ra h2t2}
    \end{subfigure}
    \hfill
    \begin{subfigure}[b]{0.49\textwidth}
        \includegraphics[width=\textwidth]{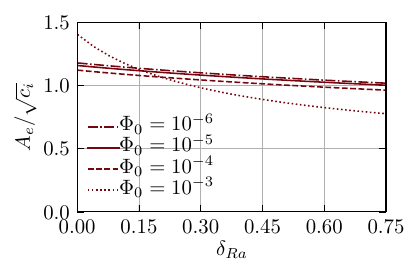}
        \caption{}
        \label{fig:Ae by sqrt(ci) vs ra h2t2}
    \end{subfigure}
    \captionsetup{width=\linewidth}
    \caption{\justifying Effect of particle volume fraction near the bifurcation point: a) variation of growth rate, b) real part of Landau constant, c) equilibrium amplitude, and d) ratio of equilibrium amplitude and the square root of growth rate with reduced Rayleigh number $\delta_{Ra}$ for other parameters kept at $\delta=0.01$, ${\Rey}_p=1$, ${R}=800$, ${E}=3385$, and $\Pran=0.71$.}
    \label{fig:bifurcations h2t2}
\end{figure}

The grid independence test on the growth rate $c_i$, the real part of the Landau constant $\Re{a_1}$, and the equilibrium amplitude $A_e$ for $\delta_{Ra}\in[0,\, 0.75]$ at three different particle volume fractions with corresponding critical wave numbers is shown in Table \ref{tab:grid independence test for nonlin-H2T2}. With an increase in the degree of the Chebychev polynomial $N$, the values of $c_i$, $\Re{a_1}$, and $A_e$ are not changed consistently for each $\Phi_0$ value. Hence, we fix $N=50$ for all the computations in the present work. Moreover, it can be seen that the growth rate $c_i$ is positive and decreases with an increase in $\Phi_0$, $\Re{a_1}<0$, and $A_e$ also decreases with an increase in $\Phi_0$. These observations are more apparent in figure \ref{fig:bifurcations h2t2}(a-c) in which $c_i$, $\Re{a_1}$, $A_e$, and $A_e/\sqrt{c_i}$ are plotted against $\delta_{Ra}$ at four different $\Phi_0$ values. Figure \ref{fig:ci vs ra h2t2} shows the linear variation of $c_i$ with $\delta_{Ra}$ near the critical point. The decrease in $c_i$ value with an increase in $\Phi_0$ reveals that the stability of the flow increases with an increase in particle volume fraction. The real part of Landau constant $\Re{a_1}<0$ for $\delta_{Ra}\in[0,\,0.75]$ at all $\Phi_0$ values (see figure \ref{fig:a1r vs ra h2t2}). Hence, $\Re{a_1}<0$ together with $c_i>0$ for $\delta_{Ra}>0$ shows the supercritical pitchfork bifurcation at the critical point. Therefore, the stable, positive and finite equilibrium amplitude $A_e$ is guaranteed \citep{drazin2004hydrodynamic,shukla2011weakly}. In the present work on particle-laden Rayleigh-B\'enard convection, the real part of the wave speed $c_r$ most dominant disturbance and the imaginary part of Landau constant $\Im{a_1}$ are always zero. Hence, from (\ref{eq:perturbation series}) and (\ref{eq:expression of A(t)}), we expect the steady-state equilibrium solutions near the critical point $\qty(\delta_{Ra}\in(0,\,0.75))$ as $t\to\infty$. Here, $c_r=\Im{a_1}=0$ can be argued physically from the symmetries of the present problem. The real part of complex wave speed $c_r$ represents the speed and the direction of propagation of disturbance in the flow. However, there is no preferential direction for the flow due to the absence of non-zero base state fluid velocity. In other words, the clockwise and counterclockwise steady rolls are the solutions to the problem, which is only possible if $c_r=0$ for the most unstable disturbance. Moreover, the present problem is invariant under translation along the horizontal direction due to its infinite extent $x\in(-\infty,\,\infty)$ and $\Im{a_1}$ represents the phase angle of amplitude (see (\ref{eq:A(t) in phase angle form})) which captures a finite spatial shift along the horizontal direction which leads to $\Im{a_1}=0$. 

The supercritical bifurcation at the critical point leads to an equilibrium amplitude $A_e$, which varies with control parameter $\delta_{Ra}$ without any hysteresis as shown in figure \ref{fig:Ae vs ra h2t2}. Since $c_i\propto\delta_{Ra}$ and from (\ref{eq:equilibrium amplitude}) it is clear that $A_e\propto\sqrt{c_i}$, hence, $A_e$ shows a square root dependency on $\delta_{Ra}$ near the critical point as shown in figure \ref{fig:Ae vs ra h2t2}. Moreover, with increase in $\Phi_0$, the growth rate $c_i$ decreases and nearly constant $\Re{a_1}$ value for $\Phi_0=\qty{10^{-6},\, 10^{-5},\, 10^{-4}}$ and decrease for $\Phi_0=10^{-3}$. Therefore, from the equation (\ref{eq:equilibrium amplitude}) for the equilibrium amplitude, $A_e$ decreases with an increase in $\Phi_0$. In the present study, $A_e\sim\sqrt{c_i}$ is the primary assumption in writing perturbation series (\ref{eq:perturbation series}) for all the dependent variables, which is shown in figure \ref{fig:Ae by sqrt(ci) vs ra h2t2}.

\begin{figure}
    \centering
    \begin{subfigure}[b]{0.49\textwidth}
        \includegraphics[width=\textwidth]{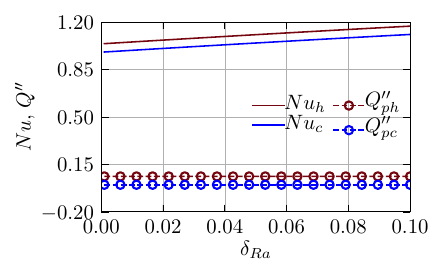}
        \caption{$\Phi_0=10^{-6}$}
        \label{fig:ra vs heat flux phi 1e-6 h2t2} 
    \end{subfigure}
    \hfill
    \begin{subfigure}[b]{0.49\textwidth}
        \includegraphics[width=\textwidth]{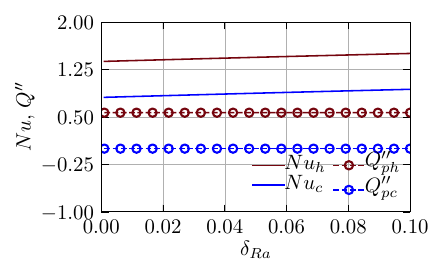}
        \caption{$\Phi_0=10^{-5}$}
        \label{fig:ra vs heat flux phi 1e-5 h2t2}
    \end{subfigure}
    \hfill
    \begin{subfigure}[b]{0.49\textwidth}
        \includegraphics[width=\textwidth]{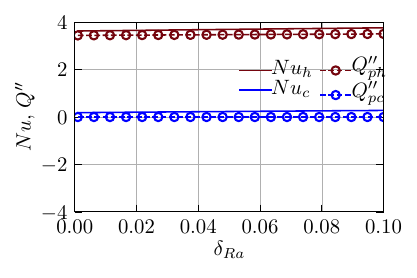}
        \caption{$\Phi_0=10^{-4}$}
        \label{fig:ra vs heat flux phi 1e-4 h2t2}
    \end{subfigure}
    \hfill
    \begin{subfigure}[b]{0.49\textwidth}
        \includegraphics[width=\textwidth]{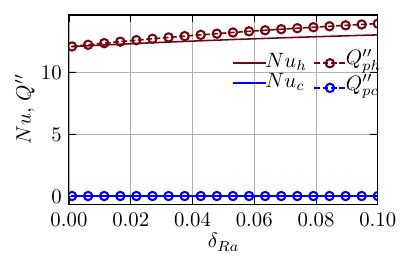}
        \caption{$\Phi_0=10^{-3}$}
        \label{fig:ra vs heat flux phi 1e-3 h2t2}
    \end{subfigure}
    \captionsetup{width=\linewidth}
    \caption{\justifying Effect of particle volume fraction on heat transfer near the bifurcation point for H2--T2 with other parameters kept constant at $\delta=0.01$, ${\Rey}_p=1$, ${R}=800$, ${E}=3385$, $\Theta_{pt}=0$ and $\Pran=0.71$. Here, continuous lines represent the fluid, and dashed lines represent the particles.}
    \label{fig:ra vs heat flux h2t2}
\end{figure}

The effect of particle volume fraction $\Phi_0$ on the Nusselt number $Nu$ and sensible heat flux $Q^{\prime\prime}_p$ by a particle at the top and bottom surface are shown in figure \ref{fig:ra vs heat flux h2t2}. The Nusselt numbers at the hot ${Nu}_h$ and cold ${Nu}_c$ surfaces are unequal due to the thermal energy coupling between the particles and fluid. As shown in (\ref{eq:net thermal energy flux balance}), the difference between ${Nu}_h$ and ${Nu}_c$ is balanced by the convective sensible heat flux by the particles at hot  ($Q^{\prime\prime}_{ph}$) and cold ( $Q^{\prime\prime}_{pc}$) surfaces. Here, the particles are introduced into the domain at the cold surface temperature ($\Theta_{pt}=0$). Hence, sensible heat flux by particles at the cold surface is $Q^{\prime\prime}_{pc}=0$. As the particle concentration increases, the difference between the heat fluxes at the hot and cold surfaces for fluid and particles increases. Thus, the Nusselt number at the hot surface increases, while at the cold surface, it tends to zero when the particle concentration increases. Physically, it means that the particles absorb all the heat flux emerging from the hot surface while settling down under gravity.

\subsection{Effect of particle injection temperature}
\begin{figure}
    \centering
    \begin{subfigure}[b]{0.49\textwidth}
        \includegraphics[width=\textwidth]{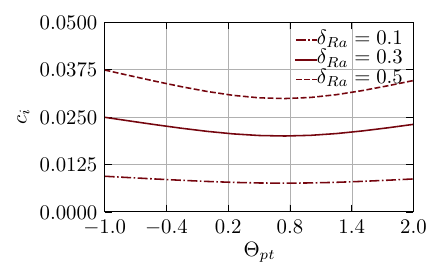}
        \caption{}
        \label{fig:tps vs ci h2t2} 
    \end{subfigure}
    \hfill
    \begin{subfigure}[b]{0.49\textwidth}
        \includegraphics[width=\textwidth]{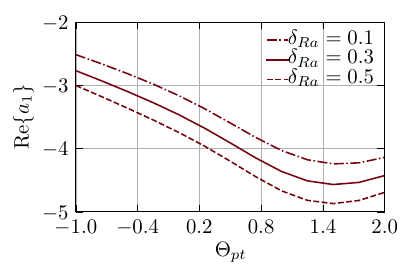}
        \caption{}
        \label{fig:tps vs a1r h2t2}
    \end{subfigure}
    \hfill
    \begin{subfigure}[b]{0.49\textwidth}
        \includegraphics[width=\textwidth]{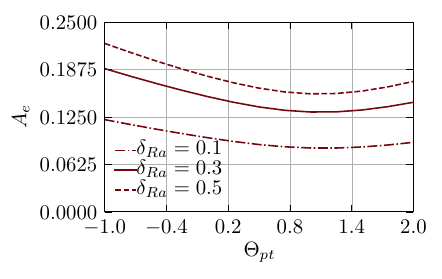}
        \caption{}
        \label{fig:tps vs Ae h2t2}
    \end{subfigure}
    \hfill
    \begin{subfigure}[b]{0.49\textwidth}
        \includegraphics[width=\textwidth]{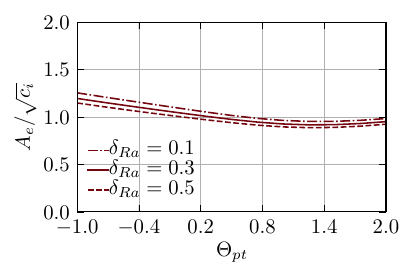}
        \caption{}
        \label{fig:tps vs Ae by sqrt(ci) h2t2}
    \end{subfigure}
    \captionsetup{width=\linewidth}
    \caption{\justifying Effect of particle initial temperature $\Theta_{pt}$ at $\delta_{Ra}=0.1$ with H2--T2 coupling: a) variation of growth rate $c_i$, b) variation of real part of Landau constant $\Re{a_1}$, c) variation of equilibrium amplitude $A_e$, and d) variation of the ratio of equilibrium amplitude and the square root of the growth rate. The other parameters are kept at $\delta=0.01$, ${\Rey}_p=1$, ${R}=800$, ${E}=3385$, $\Pran=0.71$, and $\Phi_0=10^{-4}$.}
    \label{fig:tps vs ci, a1r, Ae h2t2}
\end{figure}

Figure \ref{fig:tps vs ci, a1r, Ae h2t2} depicts the effect of particles injection temperature $\Theta_{pt}$ on the growth rate $c_i$, real part of Landau constant $\Re{a_1}$, equilibrium amplitude $A_e$, and the ratio of equilibrium amplitude and the square root of growth rate at a fixed particle volume fraction $\qty(\Phi_0=10^{-4})$ and other parameters kept at $\delta=0.01$, ${\Rey}_p=1$, ${R}=800$, ${E}=3385$, and $\Pran=0.71$. Similar to the non-monotonic effect of $\Theta_{pt}$ on the onset of instability (see figure \ref{fig:tps vs rac and kc h2t2}), $\Theta_{pt}$ shows the non-monotonic effect on the growth rate, the real part of Landau constant, and the equilibrium amplitude. As described in figure \ref{fig:tps vs rac and kc h2t2}, an increase in $\Theta_{pt}$ favours stability in the initial part, and the further increase in $\Theta_{pt}$ causes flow to favour instability. Hence, it explains the decrease in $c_i$ in the initial part and the increase in $c_i$ in the remaining part of the curves in figure \ref{fig:tps vs ci h2t2}. Figure \ref{fig:tps vs a1r h2t2} shows that for all $\Theta_{pt}\in(-1,\,2)$, $\Re{a_1}<0$ and together with $c_i>0$, the flow exhibits a supercritical pitchfork bifurcation. The equilibrium amplitude $A_e$ has a non-monotonic variation with respect to $\Theta_{pt}$, that as $\Theta_{pt}$ increase from -1, $A_e$ decreases and reaches a minimum around 1 and again increases. Finally, figure \ref{fig:tps vs Ae by sqrt(ci) h2t2} shows $A_e\sim\sqrt{c_i}$ which is the key underlying assumption while deriving the perturbation series of the form given by equation (\ref{eq:perturbation series}).

\begin{figure}
    \centering
    \includegraphics[width=0.5\linewidth]{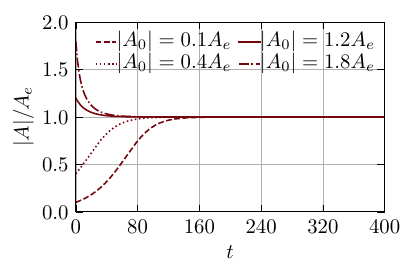}
    \captionsetup{width=\linewidth}
    \caption{\justifying Time history of amplitude function $\abs{A}$ at $\delta_{Ra}=0.1$ with different initial amplitudes $\abs{A_0}$ at $R=800$, $E=3385$, $\Phi_0=10^{-4}$, $\Theta_{pt}=0$, $\Rey_p=1$ and $\Pran=0.701$. Here, the amplitude $\abs{A}$ is normalized by the equilibrium amplitude $A_e$.}
    \label{fig:A/Ae vs time rbc}
\end{figure}

\begin{figure}
    \centering
    \begin{subfigure}[b]{0.49\textwidth}
        \includegraphics[width=\textwidth]{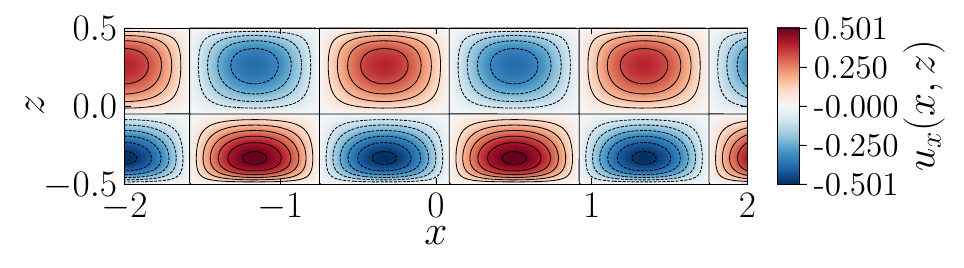}
        \caption{}
        \label{fig:linear ux secondary pattern h2t2 delta 0.1} 
    \end{subfigure}
    \hfill
    \begin{subfigure}[b]{0.49\textwidth}
        \includegraphics[width=\textwidth]{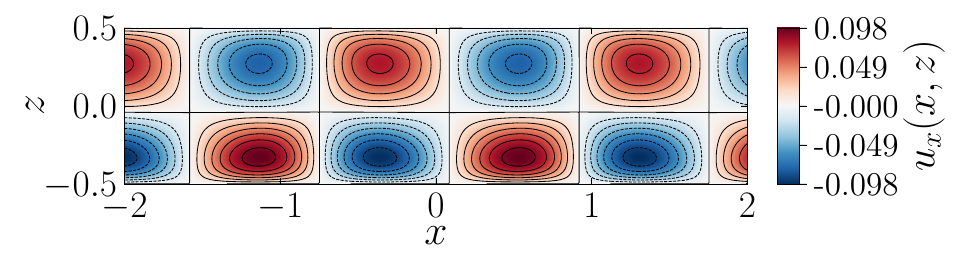}
        \caption{}
        \label{fig:non-linear ux secondary pattern h2t2 delta 0.1}
    \end{subfigure}
    \hfill
    \begin{subfigure}[b]{0.49\textwidth}
        \includegraphics[width=\textwidth]{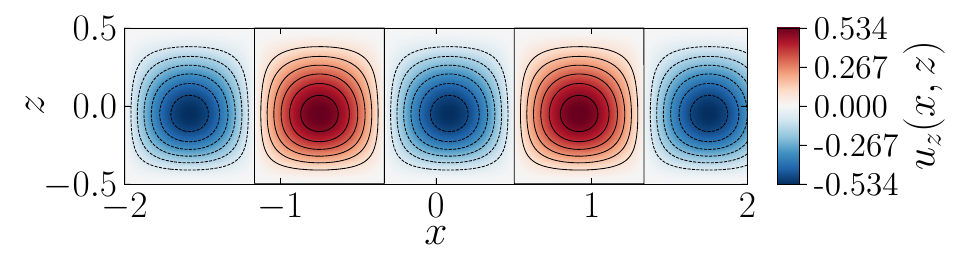}
        \caption{}
        \label{fig:linear uz secondary pattern h2t2 delta 0.1}
    \end{subfigure}
    \hfill
    \begin{subfigure}[b]{0.49\textwidth}
        \includegraphics[width=\textwidth]{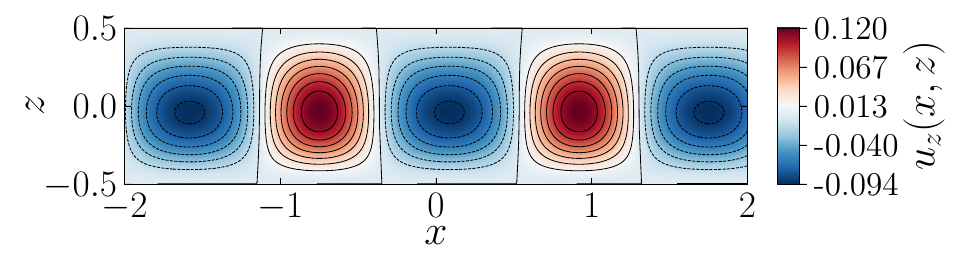}
        \caption{}
        \label{fig:non-linear uz secondary pattern h2t2 delta 0.1}
    \end{subfigure}
    \hfill
    \begin{subfigure}[b]{0.49\textwidth}
        \includegraphics[width=\textwidth]{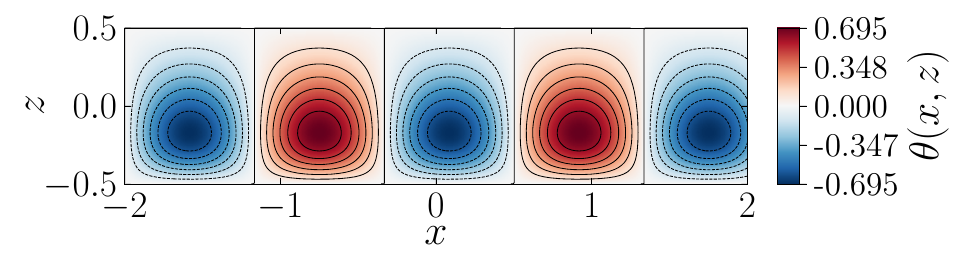}
        \caption{}
        \label{fig:linear theta secondary pattern h2t2 delta 0.1}
    \end{subfigure}
    \hfill
    \begin{subfigure}[b]{0.49\textwidth}
        \includegraphics[width=\textwidth]{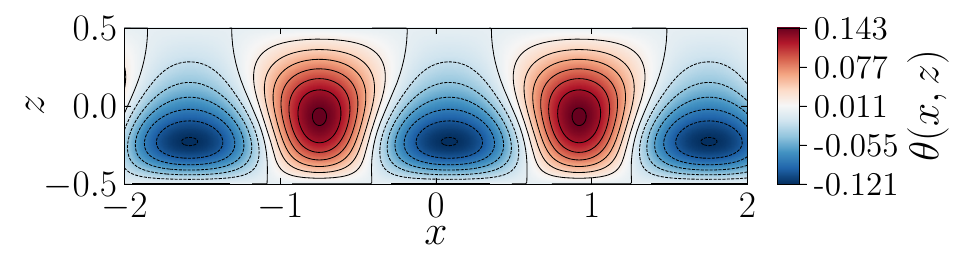}
        \caption{}
        \label{fig:non-linear theta secondary pattern h2t2 delta 0.1}
    \end{subfigure}
    \hfill
    \begin{subfigure}[b]{0.49\textwidth}
        \includegraphics[width=\textwidth]{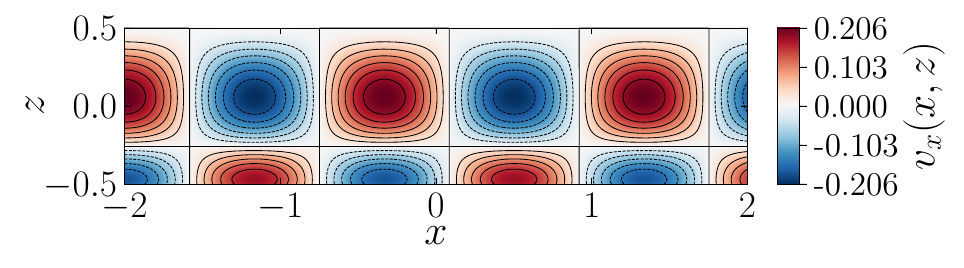}
        \caption{}
        \label{fig:linear vx secondary pattern h2t2 delta 0.1}
    \end{subfigure}
    \hfill
    \begin{subfigure}[b]{0.49\textwidth}
        \includegraphics[width=\textwidth]{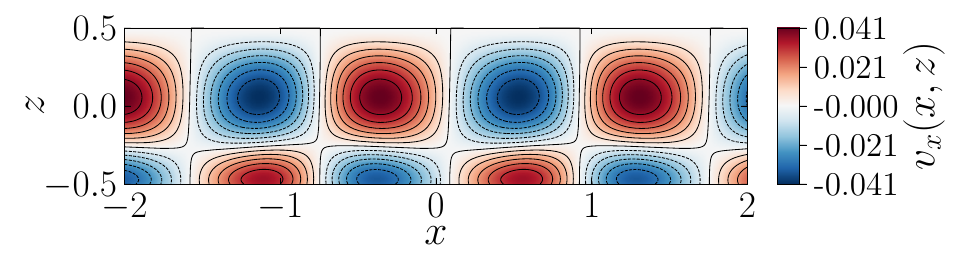}
        \caption{}
        \label{fig:non-linear vx secondary pattern h2t2 delta 0.1}
    \end{subfigure}
    \hfill
    \begin{subfigure}[b]{0.49\textwidth}
        \includegraphics[width=\textwidth]{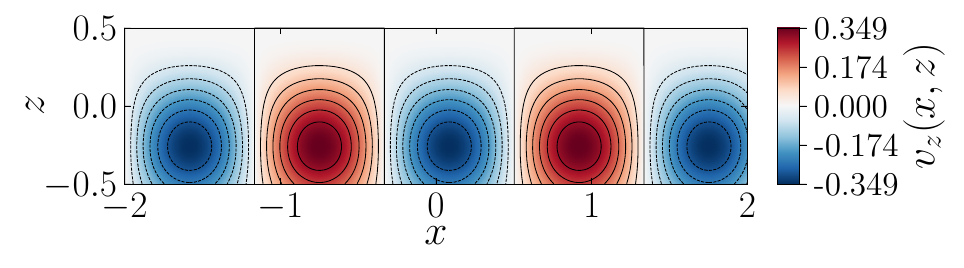}
        \caption{}
        \label{fig:linear vz secondary pattern h2t2 delta 0.1}
    \end{subfigure}
    \hfill
    \begin{subfigure}[b]{0.49\textwidth}
        \includegraphics[width=\textwidth]{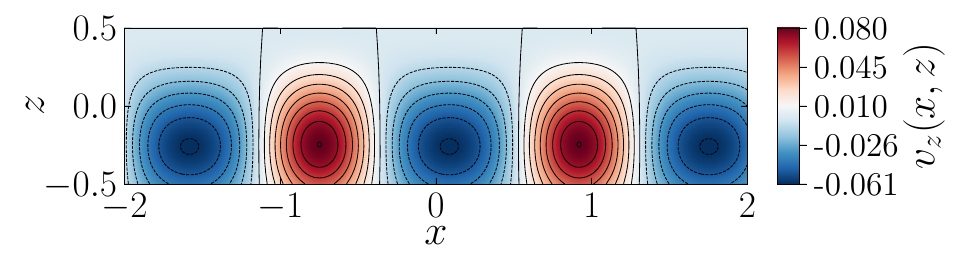}
        \caption{}
        \label{fig:non-linear vz secondary pattern h2t2 delta 0.1}
    \end{subfigure}
    \hfill
    \begin{subfigure}[b]{0.49\textwidth}
        \includegraphics[width=\textwidth]{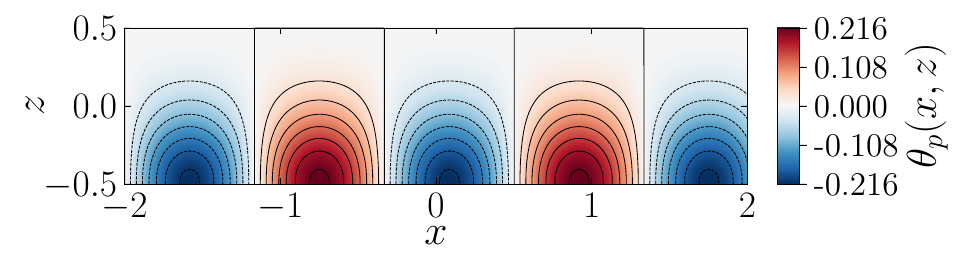}
        \caption{}
        \label{fig:linear thetap secondary pattern h2t2 delta 0.1}
    \end{subfigure}
    \hfill
    \begin{subfigure}[b]{0.49\textwidth}
        \includegraphics[width=\textwidth]{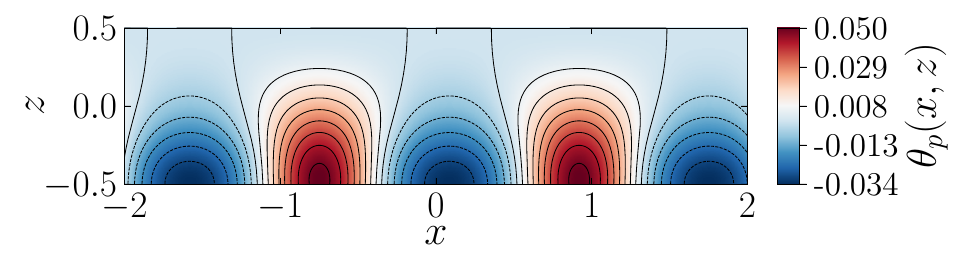}
        \caption{}
        \label{fig:non-linear thetap secondary pattern h2t2 delta 0.1}
    \end{subfigure}
    \hfill
    \begin{subfigure}[b]{0.49\textwidth}
        \includegraphics[width=\textwidth]{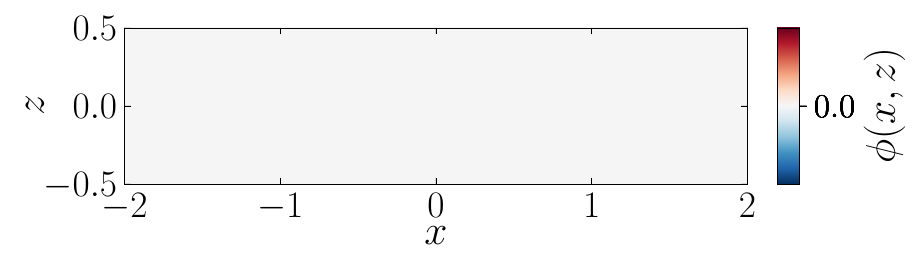}
        \caption{}
        \label{fig:linear phi secondary pattern h2t2 delta 0.1}
    \end{subfigure}
    \hfill
    \begin{subfigure}[b]{0.49\textwidth}
        \includegraphics[width=\textwidth]{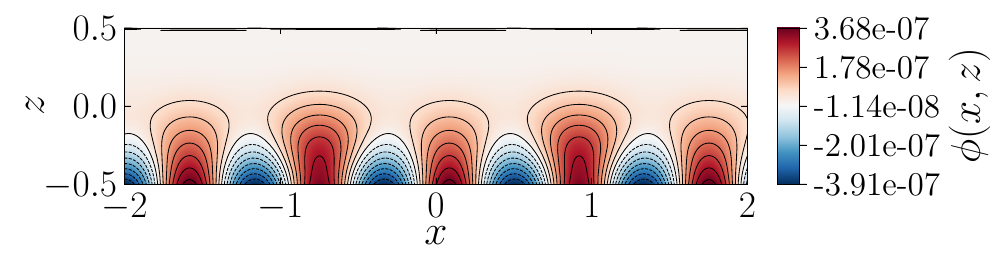}
        \caption{}
        \label{fig:non-linear phi secondary pattern h2t2 delta 0.1}
    \end{subfigure}
    \captionsetup{width=\linewidth}
    \caption{\justifying Secondary flow pattern obtained for particle-laden Rayleigh-B\'enard convection: the left-side panel (a, c, e) shows the linear analysis whereas the right-side panel (b, d, f) shows the non-linear analysis at reduced Rayleigh number $\delta_{Ra}=0.1$, with critical wave number $k_c\approx3.77$ and other parameters ${R}=800$, ${E}=3385$, $\Phi_0=10^{-4}$, $\Theta_{pt}=0$, $\Rey_p=1$ and ${\Pran}=0.701$.}
    \label{fig:secondary flow patterns linear vs non-linear h2t2 delta 0.1}
\end{figure}
\begin{figure}
    \centering
    \begin{subfigure}[b]{0.49\textwidth}
        \includegraphics[width=\textwidth]{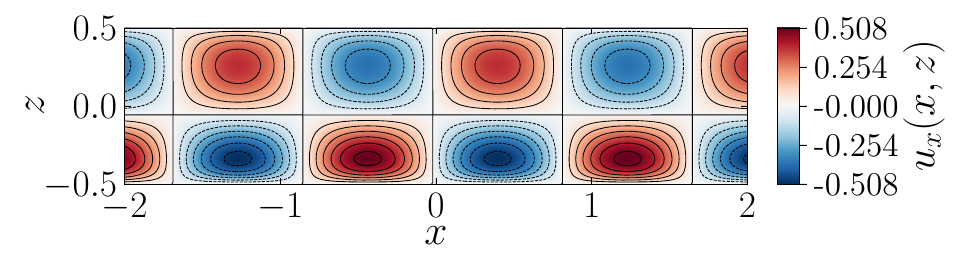}
        \caption{}
        \label{fig:linear ux secondary pattern h2t2 delta 0.5} 
    \end{subfigure}
    \hfill
    \begin{subfigure}[b]{0.49\textwidth}
        \includegraphics[width=\textwidth]{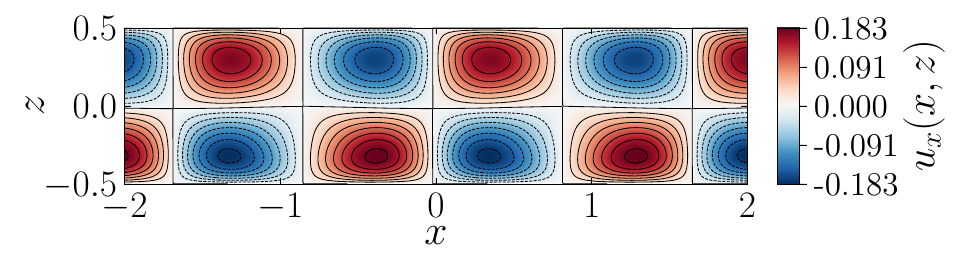}
        \caption{}
        \label{fig:non-linear ux secondary pattern h2t2 delta 0.5}
    \end{subfigure}
    \hfill
    \begin{subfigure}[b]{0.49\textwidth}
        \includegraphics[width=\textwidth]{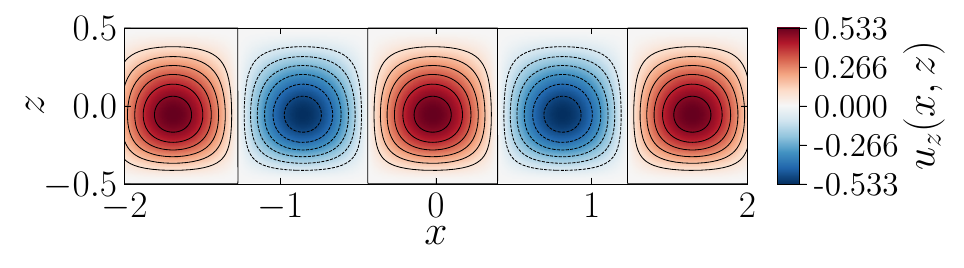}
        \caption{}
        \label{fig:linear uz secondary pattern h2t2 delta 0.5}
    \end{subfigure}
    \hfill
    \begin{subfigure}[b]{0.49\textwidth}
        \includegraphics[width=\textwidth]{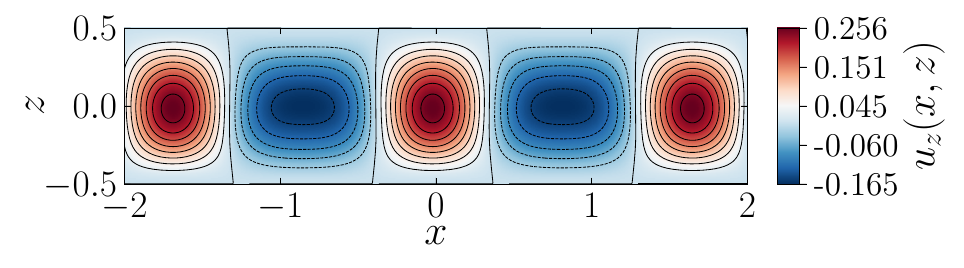}
        \caption{}
        \label{fig:non-linear uz secondary pattern h2t2 delta 0.5}
    \end{subfigure}
    \hfill
    \begin{subfigure}[b]{0.49\textwidth}
        \includegraphics[width=\textwidth]{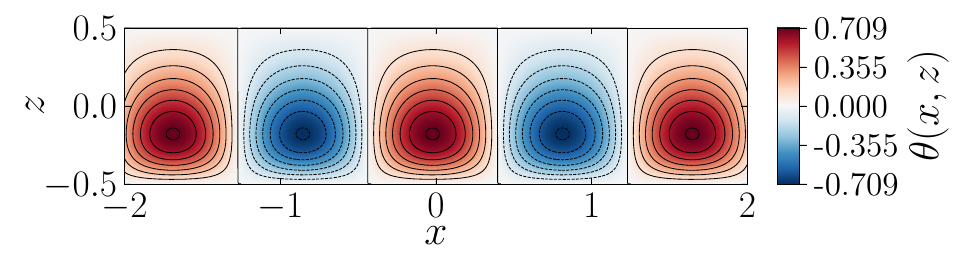}
        \caption{}
        \label{fig:linear theta secondary pattern h2t2 delta 0.5}
    \end{subfigure}
    \hfill
    \begin{subfigure}[b]{0.49\textwidth}
        \includegraphics[width=\textwidth]{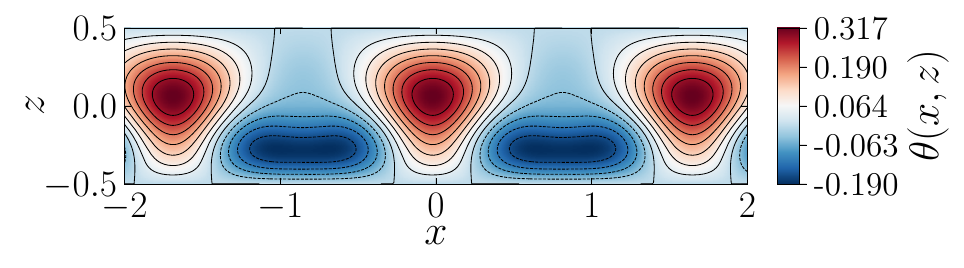}
        \caption{}
        \label{fig:non-linear theta secondary pattern h2t2 delta 0.5}
    \end{subfigure}
    \hfill
    \begin{subfigure}[b]{0.49\textwidth}
        \includegraphics[width=\textwidth]{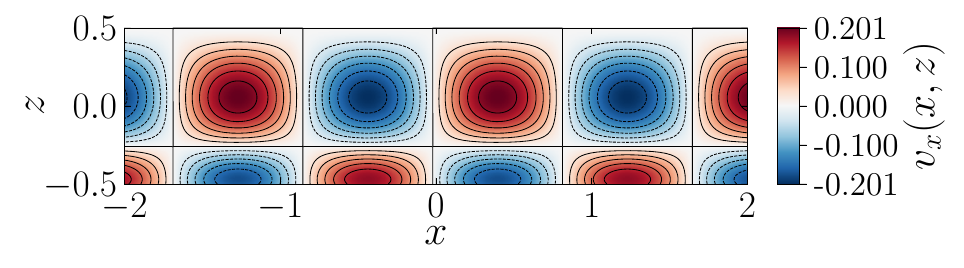}
        \caption{}
        \label{fig:linear vx secondary pattern h2t2 delta 0.5}
    \end{subfigure}
    \hfill
    \begin{subfigure}[b]{0.49\textwidth}
        \includegraphics[width=\textwidth]{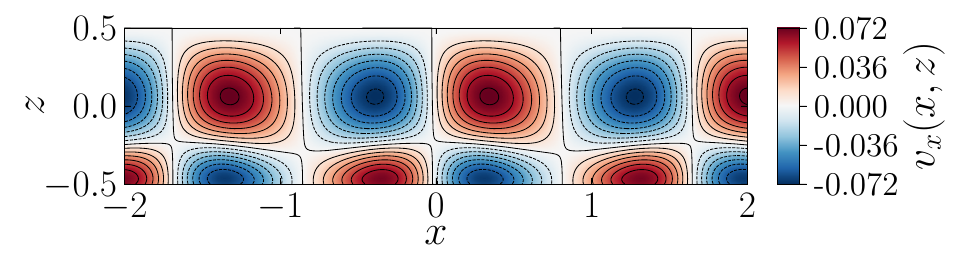}
        \caption{}
        \label{fig:non-linear vx secondary pattern h2t2 delta 0.5}
    \end{subfigure}
    \hfill
    \begin{subfigure}[b]{0.49\textwidth}
        \includegraphics[width=\textwidth]{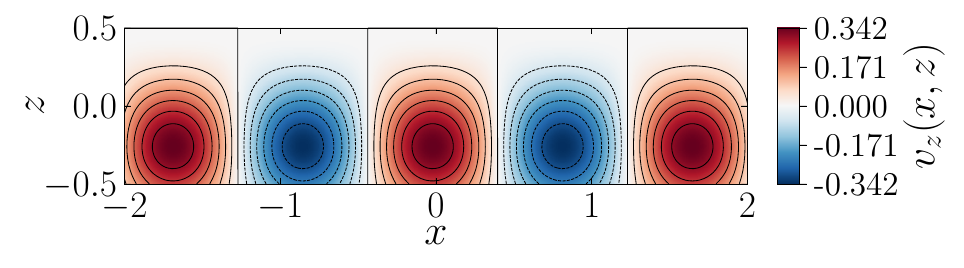}
        \caption{}
        \label{fig:linear vz secondary pattern h2t2 delta 0.5}
    \end{subfigure}
    \hfill
    \begin{subfigure}[b]{0.49\textwidth}
        \includegraphics[width=\textwidth]{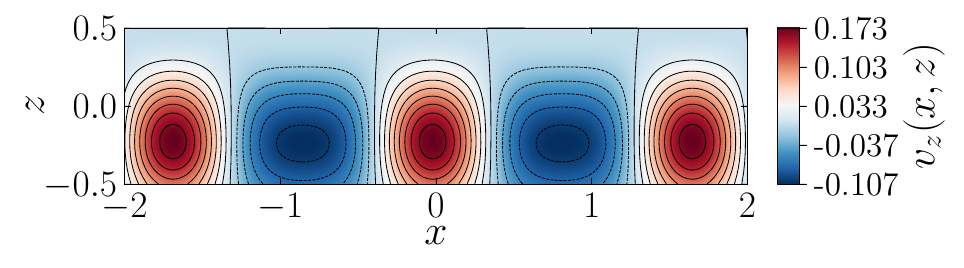}
        \caption{}
        \label{fig:non-linear vz secondary pattern h2t2 delta 0.5}
    \end{subfigure}
    \hfill
    \begin{subfigure}[b]{0.49\textwidth}
        \includegraphics[width=\textwidth]{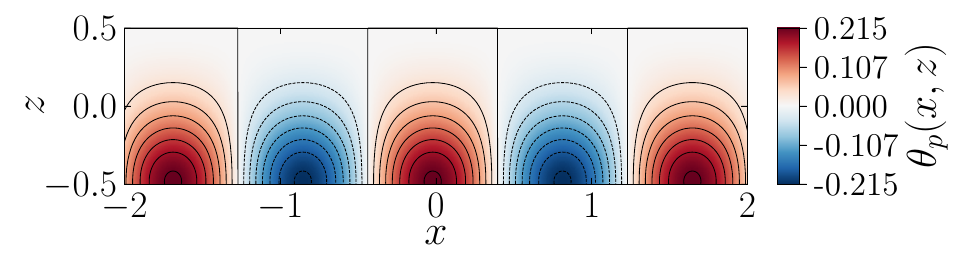}
        \caption{}
        \label{fig:linear thetap secondary pattern h2t2 delta 0.5}
    \end{subfigure}
    \hfill
    \begin{subfigure}[b]{0.49\textwidth}
        \includegraphics[width=\textwidth]{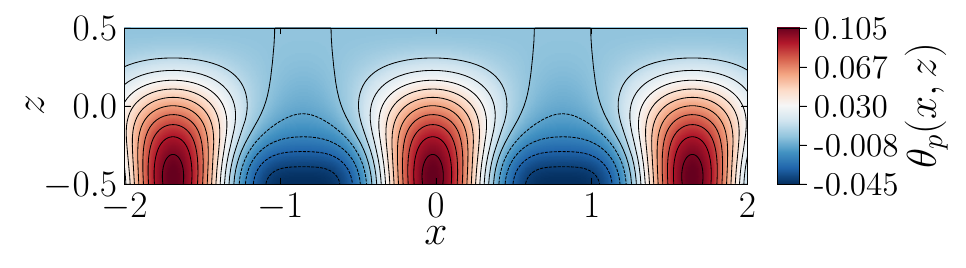}
        \caption{}
        \label{fig:non-linear thetap secondary pattern h2t2 delta 0.5}
    \end{subfigure}
    \hfill
    \begin{subfigure}[b]{0.49\textwidth}
        \includegraphics[width=\textwidth]{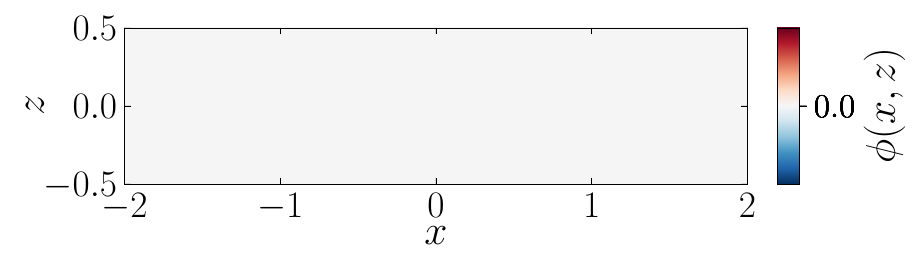}
        \caption{}
        \label{fig:linear phi secondary pattern h2t2 delta 0.5}
    \end{subfigure}
    \hfill
    \begin{subfigure}[b]{0.49\textwidth}
        \includegraphics[width=\textwidth]{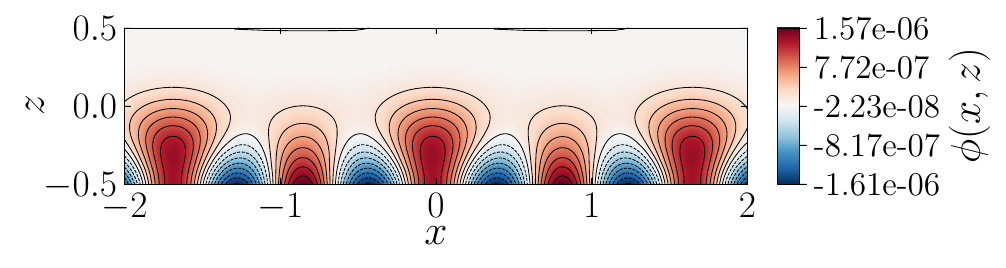}
        \caption{}
        \label{fig:non-linear phi secondary pattern h2t2 delta 0.5}
    \end{subfigure}
    \captionsetup{width=\linewidth}
    \caption{\justifying Secondary flow pattern obtained for particle-laden Rayleigh-B\'enard convection: the left-side panel (a, c, e) shows the linear analysis whereas the right-side panel (b, d, f) shows the non-linear analysis at reduced Rayleigh number $\delta_{Ra}=0.5$, with critical wave number $k_c\approx3.77$ and other parameters ${R}=800$, ${E}=3385$, $\Phi_0=10^{-4}$, $\Theta_{pt}=0$, $\Rey_p=1$ and ${\Pran}=0.701$.}
    \label{fig:secondary flow patterns linear vs non-linear h2t2 delta 0.5}
\end{figure}

\subsection{Secondary flow pattern}
Figure\ \ref{fig:A/Ae vs time rbc} shows the variation of perturbation amplitude normalized by the equilibrium amplitude with time for different initial conditions given by (\ref{eq:mod(A)^2 solution}). Due to the supercritical pitchfork bifurcation, for the given control parameter, all the perturbation amplitudes with different initial conditions tend to approach the same equilibrium amplitude as $t\to\infty$. The perturbations series given by (\ref{eq:perturbation series}) cab be rewritten for the secondary flow $\theta^\prime(x,\,z,\,t)=\theta(x,\,z,\,t)-\Theta_0(z)$ as 
\begin{equation}\label{eq:secondary flow series}
    \theta^\prime(x,\,z,\,t)=\abs{A}^2\Theta_1(z)+\mathbb{E}^1\qty{A(t)\theta_{10}(z)+A\abs{A}^2\theta_{11}(z)}+\mathbb{E}^2\qty{A^2\theta_{20}(z)}
\end{equation}
where $A(t)$ is given by (\ref{eq:expression of A(t)}). Using similar expressions for the remaining variables, we obtain the contour plots of secondary flow patterns at $\delta_{Ra}=0.1$, and $0.5$ and are shown in figures \ref{fig:secondary flow patterns linear vs non-linear h2t2 delta 0.1} and \ref{fig:secondary flow patterns linear vs non-linear h2t2 delta 0.5}, respectively. The secondary flow patterns obtained from the linear stability are shown on the left-side panel, and those obtained from the non-linear stability analysis are shown on the right-side panel. Due to the presence of particles, the top-bottom symmetry is absent even in the contours obtained from the linear stability analysis. The velocity and temperature profiles for the fluid and particles are qualitatively different due to the momentum and thermal inertia of the particles. The flow patterns are more distorted for $\delta_{Ra}=0.5$ than for $\delta_{Ra}=0.1$ due to an increase in the equilibrium amplitude with $\delta_{Ra}$. The width of the contour corresponding to the vertical component of velocity $u_z$ is larger for negative values than for positive values. Moreover, at a steady state, the net mass flux of flux across any horizontal plane must be zero to satisfy the continuity equation. Hence, the magnitude of the vertical component of velocity in upward flow regions should be higher than that in downward flow. 

From the linear stability analysis, we showed that the perturbations in particle volume fractions and divergence of particle velocity field are always zeros in the domain (see appendix \ref{appen:linear particle volume fraction solution}). However, when the perturbations are finite, the contribution from the nonlinear terms gives a non-zero distortion function $\Phi_1$, and the second harmonic function $\mathbb{E}^2$ which results in the tendency of particles to become spatially non-uniform as shown in figures \ref{fig:non-linear phi secondary pattern h2t2 delta 0.1} and \ref{fig:non-linear phi secondary pattern h2t2 delta 0.5}.

\section{Conclusions}\label{sec:conclusions}
In this work, we have performed the weakly nonlinear stability analysis of particle-laden Rayleigh-B\'enard convection. Initially, the particles are assumed to be uniformly distributed with settling velocity. At the top cold surface, new particles are injected at uniform particle volume fraction with their terminal velocity, and settled particles at the bottom hot surface are collected.  The aim of this work is to analyse the nature of bifurcation and the finite amplitude behaviour of unstable disturbances that arise outside of the linear instability boundary. We accomplished this by showing the linear stability results, and then the evolution of finite amplitude perturbation is traced using a weakly nonlinear analysis. We have analysed that, for a given particle size $\delta$, initial temperature $\Theta_{pt}$ and particle Reynolds number $\Rey_p$, the critical Rayleigh number $Ra_c$ significantly increases with an increase in particle volume fraction $\Phi_0$. However, the particle temperature $\Theta_{pt}$ shows a non-monotonic effect on the stability of the system due to its impact on the base state temperature profile.

The evolution of finite amplitude perturbation is analysed in the vicinity of the critical point. To this end, we have analysed the effect of particle volume fraction on the growth rate $c_i$, real part of Landau constant $\Re{a_1}$, and equilibrium amplitude $A_e$. It is shown that the growth rate reduces with an increase in particle volume fraction. Moreover, $\Re{a_1}>0$ for all $\Phi_0$, and together with $c_i>0$, it leads to a supercritical pitchfork bifurcation with an equilibrium amplitude $A_e=\sqrt{-k_cc_i/\Re{a_1}}$. We showed that $A_e$ reduces with increase in $\Phi_0$. We have described the significant effect of particles on the heat transfer. For instance, at steady state, the Nusselt number at the cold and hot surfaces are not equal, like in a single-phase Rayleigh-B\'enard convection. This difference in Nusselt numbers at the cold and the hot surfaces is exactly balanced by the net sensible heat flux convected by the particles. We have shown that with an increase in $\Phi_0$, the difference in the Nusselt numbers at cold and hot surfaces increases. However, we have reported that the non-monotonic effect of $\Theta_{pt}$ on $c_i$, $\Re{a_1}$, and $A_e$. Moreover, for all the range of values for $\Theta_0$ considered in the present work, the analysis showed the supercritical bifurcation for the flow. 

Finally, the variation in the secondary flow pattern due to the nonlinear interaction of different harmonics is analysed. Unlike in single-phase convection, particle-laden Rayleigh-B\'endard convection does not show top-bottom symmetry due to the directional settling of the particle under gravity, even in the linear regime. The nonlinear interaction of the fundamental modes generates Reynolds stress or distortion of the base state temperature and particle volume fraction fields. The present study reveals the local clustering of particles due to the distortion of base state particle volume fraction and higher harmonics. It is important to note that the linear stability analysis by \cite{prakhar2021linear} does not show particle clustering.

In this weakly nonlinear stability analysis, we consider nonlinear interaction only the most unstable normal mode. However, the dispersion curves obtained from the linear stability curves show the existence of a band of a continuous spectrum of unstable normal modes. The nonlinear interaction of these modes (ignored in this work) will be considered in future work, allowing the spatial modulation of the amplitude function, which satisfies the Ginzburg-Landau equation \cite{ginzburg1955theory}. Settling particles in a convection problem with phase change is more relevant to cloud microphysics. Hence, the present analysis can be easily extended to account for the effect of evaporation or condensation of the droplets in the air-water vapour mixture.

\appendix
\section{ }
\subsection{Linear disturbance equations}\label{appen:linearized disturbance equations}
Substitution of the normal modes for all the dependent variables into the linearized governing equations (\ref{eq:linear fluid masss balance})--(\ref{eq:linear partilce energy}) leads to a system of linear ordinary differential equations. The resulting equations are represented in the linear operator form as, 
\begin{eqnarray}\label{eq:linear continuity equation}
    &&\quad\mathscr{L}_0\qty(k,\,\hat{u}_x,\,\hat{u}_z)=\text{i}k\hat{u}_x+\dv{\hat{u}_z}{z}=0,\\
    &&\quad\mathscr{L}_x\qty(k,\,c,\,\hat{u}_x,\,\hat{p},\,\hat{v}_x,\,\Phi_0,\,{Ra},\,{\Pran},\,{R},\,{St}_m)\nonumber\\
    &&\qquad=\text{i}ck\hat{u}_x+\sqrt{\frac{{\Pran}}{{Ra}}}\qty(\dv[2]{z}-k^2)\hat{u}_x-\text{i}k\hat{p}-\qty(\frac{\Phi_0{R}}{{St}_m})\qty(\hat{u}_x-\hat{v}_x)=0,\\
    &&\quad\mathscr{L}_z\qty(k,\,c,\,\hat{u}_z,\,\hat{p},\,\hat{\theta},\,\hat{v}_z,\,\hat{\phi},\,v_0,\,\Phi_0,\,{Ra},\,{\Pran},\,{R},\,{St}_m)\nonumber\\
    &&\qquad=\text{i}ck\hat{u}_z+\sqrt{\frac{{\Pran}}{{Ra}}}\qty(\dv[2]{z}-k^2)\hat{u}_z-\dv{\hat{p}}{z}+\hat{\theta}-\qty(\frac{\Phi_0{R}}{{St}_m})\qty(\hat{u}_z-\hat{v}_z)-\qty(\frac{{R}v_0}{{St}_m})\hat{\phi}=0,\\
    && \quad\mathscr{L}_\theta\qty(k,\,c,\,\hat{u}_z,\,\hat{\theta},\,\hat{\theta}_p,\,\hat{\phi},\,\Theta_0,\,\Theta_{p0},\,\Phi_0,\,{Ra},\,{\Pran},\,{E},\,{St}_{th})\nonumber\\
    &&\qquad=\text{i}ck\hat{\theta}+\frac{1}{\sqrt{{RaPr}}}\qty(\dv[2]{z}-k^2)\hat{\theta}-\dv{\Theta_0}{z}\hat{u}_z-\frac{{E}\Phi_0}{{St}_{th}}\qty(\hat{\theta}-\hat{\theta_p})-\frac{{E}\qty(\Theta_0-\Theta_{p0})}{{St}_{th}}\hat{\phi}=0,\\
    && \quad\mathscr{L}_{\phi}(k,\,c,\,\hat{v}_x,\,\hat{v}_z,\,v_0,\,\Phi_0)=\text{i}ck\hat{\phi}-\Phi_0\qty(\text{i}k\hat{v}_x+\dv{\hat{v}_z}{z})+v_0\dv{\hat{\phi}}{z}=0,\\
    && \quad\mathscr{L}_{px}\qty(k,\,c,\,\hat{u}_x,\,\hat{v}_x,\,v_0,\,{St}_m)=\text{i}ck\hat{v}_x+v_0\dv{\hat{v}_x}{z}+\frac{\hat{u}_x-\hat{v}_x}{{St}_{m}}=0,\\
    && \quad\mathscr{L}_{pz}\qty(k,\,c,\,\hat{u}_z,\,\hat{v}_z,\,v_0,\,{St}_m)=\text{i}ck\hat{v}_z+v_0\dv{\hat{v}_z}{z}+\frac{\hat{u}_z-\hat{v}_z}{{St}_m}=0,\\ \label{eq:thetap linear equation}
    && \quad\mathscr{L}_{p\theta}\qty(k,\,c,\,\hat{\theta},\,\hat{v}_z,\,\hat{\theta}_p,\,v_0,\,\Theta_{p0},\,{St}_{th})=\text{i}ck\hat{\theta}_{p}+v_0\dv{\hat{\theta}_p}{z}+\frac{\hat{\theta}-\hat{\theta}_p}{{St}_{th}}-\dv{\Theta_{p0}}{z}\hat{v}_z=0
\end{eqnarray}
for $z\in\qty(-1/2,\,1/2)$.

\subsection{Scalar functions}\label{appen:scalar functions}
In the derivation of the Landau equation, the following scalar functions emerge due to the nonlinear interaction of different modes.
\begin{eqnarray}\label{eq:G functions for h2t2}
    \mathcal{G}_x&=&\Tilde{u}_{z10}\dv{u_{x20}}{z}+u_{z20}\dv{\Tilde{u}_{x10}}{z}-\text{i}k_cu_{x20}\Tilde{u}_{x10}+2u_{x20}\dv{\Tilde{u}_{z10}}{z}\nonumber\\
    &&\qquad +\frac{{R}}{{St}_m}\qty(\Phi_1\qty{u_{x10}-v_{x10}}+\phi_{20}\qty{\Tilde{u}_{x10}-\Tilde{v}_{x10}}+\Tilde{\phi}_{10}\qty{u_{x20}-v_{x20}}),\\
    \mathcal{G}_z&=&\Tilde{u}_{z10}\dv{u_{z20}}{z}+u_{z20}\dv{\Tilde{u}_{z10}}{z}-\text{i}k_cu_{x20}\Tilde{u}_{z10}+2u_{z20}\dv{\Tilde{u}_{z10}}{z}\nonumber\\
    &&\qquad +\frac{{R}}{{St}_m}\left(\Phi_1\qty{u_{z10}-v_{z10}}+\phi_{20}\qty{\Tilde{u}_{z10}-\Tilde{v}_{z10}}\right.\nonumber\\
    &&\qquad\qquad\left.+\Tilde{\phi}_{10}\qty{u_{z20}-v_{z20}}-\phi_{10}V_1\right),\\
    \mathcal{G}_\theta&=&\Tilde{u}_{z10}\dv{\theta_{20}}{z}+u_{z20}\dv{\Tilde{\theta}_{10}}{z}-\text{i}k_cu_{x20}\Tilde{\theta}_{10}+2\theta_{20}\dv{\Tilde{u}_{z10}}{z}+u_{z10}\dv{\Theta_1}{z}\nonumber\\
    &&\qquad+\frac{{E}}{{St}_{th}}\left(\phi_{10}\qty{\Theta_1-\Theta_{p1}}+\Tilde{\phi}_{10}\qty{\theta_{20}-\theta_{p20}}\right.\nonumber\\
    &&\qquad\qquad\left.+\Phi_1\qty{\theta_{10}-\theta_{p10}}+\phi_{20}\qty{\Tilde{\theta}_{10}-\Tilde{\theta}_{p10}}\right)\\
    \mathcal{G}_\phi&=&\dv{z}\qty(\Tilde{v}_{z10}\phi_{20})+\dv{z}\qty(v_{z20}\Tilde{\phi}_{10})+\text{i}k_c\qty( v_{x20}\Tilde{\phi}_{10}+\Tilde{v}_{x10}\phi_{20}+\Phi_1v_{x10})\nonumber\\
    &&\qquad+\dv{z}\qty(\Phi_1v_{z10})+\dv{z}\qty(\phi_{10}V_1),\\
    \mathcal{G}_{px}&=&\Tilde{v}_{z10}\dv{v_{x20}}{z}+v_{z20}\dv{\Tilde{v}_{x10}}{z}-\text{i}k_cv_{x20}\Tilde{v}_{x10}+2\text{i}k_c\Tilde{v}_{x10}v_{x20}+V_1\dv{v_{x10}}{z},\\
    \mathcal{G}_{pz}&=&\Tilde{v}_{z10}\dv{v_{z20}}{z}+v_{z20}\dv{\Tilde{v}_{z10}}{z}-\text{i}k_cv_{x20}\Tilde{v}_{z10}+2\text{i}k_c\Tilde{v}_{x10}v_{z20}+\dv{z}\qty(V_1v_{z10}),\\
    \mathcal{G}_{p\theta}&=&\Tilde{v}_{z10}\dv{{\theta}_{p20}}{z}+v_{z20}\dv{\Tilde{\theta}_{p10}}{z}-\text{i}k_cv_{x20}\Tilde{\theta}_{p10}+2\text{i}k_c\Tilde{v}_{x10}\theta_{p20}\nonumber\\
    &&\qquad+V_1\dv{\theta_{p10}}{z}+v_{z10}\dv{\Theta_{p1}}{z}.
\end{eqnarray}

\subsection{Linear adjoint equations}\label{appen:linear adjoint equations}
Adjoint of the linear operator $\mathscr{L}$ is defined as
\begin{equation}\label{eq:adjoint def.}
    \ev{\vb{X}^\dag,\,\mathscr{L}\vb{X}}=\ev{\mathscr{L}^\dag\vb{X}^\dag,\,\vb{X}},
\end{equation}
where $\vb{X}^\dag$ is the eigenvector corresponding to the adjoint operator $\mathscr{L}^\dag$. The definition for the inner product used in (\ref{eq:adjoint def.}) for the two real-valued vector functions $\vb{X}(z)=\mqty[x_1&x_2&x_3&x_4&x_5&x_6&x_7&x_8]^T$ and $\vb{Y}(z)=\mqty[y_1&y_2&y_3&y_4&y_5&y_6&y_7&y_8]^T$ is given by
\begin{equation}\label{eq:inner product def.}
    \ev{\vb{X},\,\vb{Y}}=\displaystyle\int_{z=-1/2}^{1/2} \left[\vb{X}\right]^T\vb{Y} \,dz=\int_{z=-1/2}^{1/2} \sum_{j=1}^8 {x}_j(z)y_j(z)\,dz.
\end{equation}
Using the definition for adjoint in (\ref{eq:adjoint def.}), the adjoint linear system corresponding to the linear stability equations (\ref{eq:linear continuity equation})--(\ref{eq:thetap linear equation}) is given by
\begin{eqnarray}
        &&\quad\mathscr{L}_0^\dag(k_c,\,\hat{u}_x^\dag,\,\hat{u}_z^\dag)=-\text{i}k_c\hat{u}_x^\dag+\dv{\hat{u}_z^\dag}{z}=0,\\
        &&\quad\mathscr{L}_x^\dag\qty(k_c,\,c,\,u_x^\dag,\,\hat{p}^\dag,\,\hat{v}_x^\dag,\,\Phi_0,\,{Ra},\,{\Pran},\,{R},\,{St}_m)=\text{i}ck_c\hat{u}_x^\dag+\sqrt{\frac{{\Pran}}{{Ra}}}\qty(\dv[2]{z}-k_c^2)\hat{u}_x^\dag+\text{i}k_c\hat{p}^\dag\nonumber\\
        &&\qquad\qquad -\qty(\frac{\Phi_0{R}}{{St}_m})\hat{u}_x^\dag+\frac{\hat{v}_x^\dag}{{St}_m}=0,\\
        &&\quad\mathscr{L}_z^\dag\qty(k_c,\,c,\,u_z^\dag,\,\hat{p}^\dag,\,\hat{\theta}^\dag,\,\hat{v}_z^\dag,\,\Theta_0,\,\Phi_0,\,{Ra},\,{\Pran},\,{R},\,{St}_m)=\text{i}ck_c\hat{u}_z^\dag+\sqrt{\frac{{\Pran}}{{Ra}}}\qty(\dv[2]{z}-k_c^2)\hat{u}_z^\dag\nonumber\\
        &&\qquad\qquad-\dv{\hat{p}}{z}^\dag-\dv{\Theta_0}{z}\hat{\theta}^\dag-\qty(\frac{\Phi_0{R}}{{St}_m})\hat{u}_z^\dag+\frac{\hat{v}_z^\dag}{{St}_m}=0,\\
        &&\quad\mathscr{L}_\theta^\dag\qty(k_c,\,c,\,\hat{u}_z^\dag,\,\hat{\theta}^\dag,\,\hat{\theta}_p^\dag,\,\Phi_0,\,{Ra},\,{\Pran},\,{St}_{th},\,{E})=\text{i}ck_c\hat{\theta}^\dag+\frac{1}{\sqrt{{Ra\Pran}}}\qty(\dv[2]{z}-k_c^2)\hat{\theta}^\dag+\hat{u}_z^\dag \nonumber\\
        &&\qquad\qquad-\frac{{E}\Phi_0}{{St}_{th}}\hat{\theta}^\dag +\frac{\hat{\theta}_p^\dag}{{St}_{th}}=0,\\
        &&\quad\mathscr{L}_{\phi}^\dag\qty(k_c,\,c,\,\hat{u}_z^\dag,\,\hat{\theta}^\dag,\,\hat{\phi}^\dag,\,\Theta_0,\,v_0,\,\Theta_{p0},\,{St}_m,\,{R},\,{St}_{th},\,{E})=\text{i}ck_c\hat{\phi}^\dag-v_0\dv{\hat{\phi}^\dag}{z}-\frac{{R}v_0}{{St}_m}\hat{u}_z^\dag\nonumber\\
        &&\qquad\qquad-\frac{{E}\qty(\Theta_0-\Theta_{p0})}{{St}_{th}}\hat{\theta}^\dag=0,\\
        &&\quad\mathscr{L}_{px}^\dag\qty(k_c,\,c,\,\hat{u}_x^\dag,\,\hat{v}_x^\dag,\,\hat{\phi}^\dag,\,v_0,\,\Phi_0,\,{R},\,{St}_m)=\text{i}ck_c\hat{v}_x^\dag-v_0\dv{\hat{v}_x^\dag}{z}-\frac{\hat{v}_x^\dag}{{St}_m}-\text{i}k_c\Phi_0\hat{\phi}^\dag\nonumber\\
        &&\qquad\qquad+\frac{\Phi_0{R}}{{St}_m}\hat{u}_x^\dag=0,\\
        &&\quad\mathscr{L}_{pz}^\dag\qty(k_c,\,c,\,\hat{u}_z^\dag,\,\hat{v}_z^\dag,\,\hat{\theta}_p^\dag,\,\hat{\phi}^\dag,\,\Theta_0,\,v_0,\,\Phi_0,\,{R},\,{St}_m)=\text{i}ck_c\hat{v}_z^\dag-v_0\dv{\hat{v}_z^\dag}{z}-\frac{\hat{v}_z^\dag}{{St}_m}+\Phi_0\dv{\hat{\phi}^\dag}{z}\nonumber\\
        &&\qquad\qquad+\frac{\Phi_0{R}}{{St}_m}\hat{u}_z^\dag-\dv{\Theta_{p0}}{z}\hat{\theta}_p^\dag=0,\\
        &&\quad\mathscr{L}_{p\theta}\qty(k_c,\,c,\,\hat{\theta}^\dag,\,\hat{\theta}_p^\dag,\,v_0,\,\Phi_0,\,{E},\,{St}_{th})=\text{i}ck_c\hat{\theta}_p^\dag-v_0\dv{\hat{\theta}_p^\dag}{z}-\frac{\hat{\theta}_p^\dag}{{St}_{th}}+\frac{{E}\Phi_0}{{St}_{th}}\hat{\theta}^\dag=0,
\end{eqnarray}
for $z\in\qty(-1/2,\,1/2)$ with boundary conditions given by
\begin{eqnarray}
    \text{At }z&=&-1/2:\text{ }\hat{u}_x^\dag=\hat{u}_z^\dag=\hat{\theta}^\dag=\hat{v}_x^\dag=\hat{v}_z^\dag=\hat{\theta}_p^\dag=\hat{\phi}^\dag=0,\\
    \text{At }z&=&1/2:\text{ }\hat{u}_x^\dag=\hat{u}_z^\dag=\hat{\theta}^\dag=0.
\end{eqnarray}
The adjoint eigenfunctions are normalized such that
\begin{equation}
\int_{-1/2}^{1/2}\qty(u_{x10}\hat{u}_x^\dag+u_{z10}\hat{u}_z^\dag+\theta_{10}\hat{\theta}^\dag+\phi_{10}\hat{\phi}^\dag+v_{x10}\hat{v}_x^\dag+v_{z10}\hat{v}_z^\dag+\theta_{p10}\hat{\theta}_p^\dag)\,dz=1.
\end{equation}

\subsection{Numerical procedure}\label{appen:numerical procedure}
The basic state equations are solved analytically, and the linear and nonlinear stability equations in the present work are solved using the spectral Chebychev collocation method \citep{boyd2001chebyshev}. The underlying system of linear and non-linear equations, along with the boundary conditions, is transformed into the Chebychev polynomial domain, that is, $\qty[-1,\,1]$ using the transformation $\xi=2z$. Here, the continuous variable $\xi$ discretized onto a collocated Gauss-Labatto points, $\xi_j$ given by
\begin{equation}
    \xi_j=\cos\qty(\frac{(j-1)\pi}{{N}-1}),\qfor j={1, \,2, \,..., \,N},
\end{equation}
where ${N}-1$ is the degree of the Chebychev polynomial. All derivatives are obtained using the MATLAB differential matrix suite by \cite{weideman2000matlab}. The generalized eigenvalue problem of the form $\mathbb{A}_d\vb{q}_d=c\mathbb{B}_d\vb{q}_d$ is solved using QZ-algorithm \citep{moler1973algorithm} in MATLAB using \texttt{eig} command for the linear stability analysis. Here, $\vb{q}_d$ is a discretized eigenvector, $\mathbb{A}_d$, and $\mathbb{B}_d$ are the discretized complex square matrices. The system of adjoint equations of the linear stability problem is solved using a similar spectral method. In nonlinear stability analysis, we need to solve a system of nonhomogeneous equations of the form $\mathcal{A}{X}=b$, here $\mathcal{A}=\mathbb{A}_d-c\mathbb{B}_d$. For the harmonic $\mathbb{E}^2$, the wave number $k_c$ is replaced by $2k_c$, and the complex wave-speed $c$ of the fundamental harmonic $\mathbb{E}^1$ is not an eigenvalue of this system. Hence, the matrix $\mathcal{A}=\mathbb{A}_d-c\mathbb{B}_d$ is non-singular and can be inverted and gives a unique solution for the harmonic $\mathbb{E}^2$. Similarly, the distortion functions also result in a system of non-singular, non-homogeneous equations, which are solved in a similar manner. For the harmonic $\mathbb{E}^1$ at $\order{c_i^{3/2}}$, the system $\mathcal{A}X_{11}=b$ turns out to be singular, and the existence of the solution is guaranteed by applying the Fredholm alternative. However, the solution $X_{11}$ is non-unique; hence, one of the solutions is obtained using the Singular Value Decomposition (SVD) built into the MATLAB software. All the integrals required for obtaining Landau constant $a_1$ and normalization are evaluated using the Gauss-Chebychev quadrature formula.

The numerical code developed for solving the linear stability equations is validated by comparing the results with earlier work by \cite{prakhar2021linear}. The results generated by the code are in excellent agreement with the published results. For the nonlinear stability calculations, the results remain consistent when the order of the polynomial $(N)$ is 50 or more (shown in table \ref{tab:grid independence test for nonlin-H2T2}). Therefore, for all the computations, the polynomial order is taken to be 50.

\subsection{Solution to linearized particle concentration}\label{appen:linear particle volume fraction solution}
From the above, the linearized equations for the particle volume fraction and the divergence of the particle velocity field for the particle-laden Rayleigh-B\'enard convection are given by
\begin{eqnarray}\label{eq:phi' equation}
        \pdv{\phi'}{t}&=&v_0\pdv{\phi'}{z}-\Phi_0\qty(\pdv{v_x'}{x}+\pdv{v_z'}{z}),\\ \label{eq:divv equation}
        \pdv{t}\qty(\pdv{v_x'}{x}+\pdv{v_z'}{z})&=&v_0\pdv{z}\qty(\pdv{v_x'}{x}+\pdv{v_z'}{z})-\frac{1}{{St}_m}\qty(\pdv{v_x'}{x}+\pdv{v_z'}{z})
\end{eqnarray}
for $\qty(x,\,z,\,t)\in\qty{\mathbb{R}\times(-1/2,\,1/2)\times\qty(0,\,\infty)}$.
Since the above equations are linear and coefficients are only functions of spatial variables $(x,\,z)$, the general solution can be expressed as
\begin{eqnarray}\label{eq:phi' solution form}
    \phi'&=&f(x,\,z)e^{\alpha t},\\ \label{eq:divv solution form}
    \pdv{v_x'}{x}+\pdv{v_z'}{z}&=&g(x,\,z)e^{\alpha t},
\end{eqnarray}
where $\alpha\in\mathbb{C}$ in general. By substituting (\ref{eq:phi' solution form})--(\ref{eq:divv solution form}) in (\ref{eq:phi' equation})--(\ref{eq:divv equation}), we obtain
\begin{eqnarray}
    \qty(\pdv{v_x'}{x}+\pdv{u_z'}{z})(x,\,z,\,t)&=&f_t(x)\exp{-{\qty({\alpha}+\frac{1}{{St}_m})}\frac{\qty(1/2-z)}{v_0}}e^{\alpha t},\\
    \phi'\qty(x,\,z,\,t)&=&\qty(g_t(x)-f_t(x)\qty(\frac{\Phi_0{St}_m}{v_0})\qty[1-\exp{-\frac{1}{{St}_m}\frac{\qty(1/2-z)}{v_0}}])\nonumber\\
    &&\exp{-{\alpha}\frac{\qty(1/2-z)}{v_0}},
\end{eqnarray}
where $f_t(x)=f(x,\,1/2)$ and $g_t(x)=g(x,\,1/2)$ are the boundary conditions at the top boundary $z=1/2$. If we specify the following boundary conditions
\begin{align}
    \text{At }z=1/2:\text{ }\pdv{v_x'}{x}+\pdv{u_z'}{z}=\phi'=0\text{ for }x\in \mathbb{R},
\end{align}
then $f_t=g_t=0$, results in $\phi'=\pdv{v_x'}{x}+\pdv{v_z'}{z}=0$ for $\qty(x,\,z,\,t)\in\qty{\mathbb{R}\times(-1/2,\,1/2)\times\qty(0,\,\infty)}$. Hence, the particles initialized uniformly with concentration $\Phi_0$ remain uniform in the domain forever in the presence of infinitesimal perturbations \citep{prakhar2021linear}.

\bibliographystyle{jfm}
\bibliography{jfm}
\end{document}